\documentclass[12pt,preprint]{emulateapj}
\begin{document}

\parindent=1.0cm

\title{CHANGES: THE PAST, PRESENT, AND FUTURE OF THE NEARBY DWARF GALAXY NGC 55 \altaffilmark{1} \altaffilmark{2}}

\author{T. J. Davidge}

\affil{Dominion Astrophysical Observatory,
\\National Research Council of Canada, 5071 West Saanich Road,
\\Victoria, BC Canada V9E 2E7\\tim.davidge@nrc.ca}

\altaffiltext{1}{Based on observations obtained at the Gemini Observatory, which is
operated by the Association of Universities for Research in Astronomy, Inc., under a
cooperative agreement with the NSF on behalf of the Gemini partnership: the National
Science Foundation (United States), the National Research Council (Canada), CONICYT
(Chile), Minist\'{e}rio da Ci\^{e}ncia,
Tecnologia e Inova\c{c}\~{a}o (Brazil) and Ministerio de Ciencia, Tecnolog\'{i}a e
Innovaci\'{o}n Productiva (Argentina).}

\altaffiltext{2}{This research has made use of the NASA/IPAC Infrared Science Archive,
which is operated by the Jet Propulsion Laboratory, California Institute of Technology,
under contract with the National Aeronautics and Space Administration.}

\begin{abstract}

	Spectra that cover wavelengths from 0.6 to $1.1\mu$m are used to examine the 
behavior of emission and absorption features in a contiguous $22 \times 300$ arcsec region 
centered on the nearby dwarf galaxy NGC 55. This area includes the two largest star-forming 
complexes in the galaxy, as well as other star-forming structures in the little-explored 
north west part of the disk. Based on the relative strengths of various emission 
features measured over spatial scales of many tens of parsecs, 
it is concluded that the ionization states and sulphur abundances in most of the 
star-forming regions near the center of NGC 55 are similar. However, a large star-forming 
region is identified in the north west part of the disk at 
a projected distance of $\sim 1$ kpc from the galaxy 
center that has distinct ionization properties. In addition to tracing areas of present-day 
star formation, the spectra are also used to identify fossil star-forming regions by 
mapping the depth of the near-infrared Ca triplet. One such area 
is identified near the intersection of the major and minor axes. There is a 
corresponding concentration of bright red stars in archival [3.6] 
and [4.5] images that are part of a mass concentration 
that is structurally distinct from the surrounding disk. It is 
suggested that the area near the intersection of the major and minor axes in NGC 55 is 
a proto-nucleus. The spectra of bright unresolved sources that are 
blended stellar asterisms, compact HII regions, and star clusters are also discussed. 
The spectra of some of the HII regions contain Ca triplet absorption lines, signalling 
a concentration of stars in the resolution element that span many Myr. Six of 
the unresolved sources have spectroscopic characteristics that are 
indicative of C stars embedded in intermediate age clusters, and these are likely 
compact star clusters that are analogous to those in the Large 
Magellanic Cloud. The peculiar properties of NGC 55 have been well 
documented in the literature, and it is argued that 
these may indicate that NGC 55 is transforming into a dwarf lenticular galaxy.

\end{abstract}

\keywords{galaxies:individual (NGC 55) -- galaxies:ISM -- galaxies:stellar content}

\section{INTRODUCTION}

	Studies of nearby galaxies lay the foundation for understanding 
the evolution of more distant systems. Given the wide diversity of galaxy types 
throughout the Universe, it is then important to examine the evolution of as many 
nearby galaxies as possible. At a distance of $\sim 2.3$ Mpc (Kudritzki et al. 2016 
and references therein), NGC 55 is one of the nearest galaxies 
that is not a member of the Local Group. However, many of its properties remain 
uncertain, in large part because it is viewed with an almost edge-on orientation. 

	While NGC 55 has been classified as a Magellanic Irregular (de Vaucouleurs et 
al. 1991), the distribution of light at red and near-infrared wavelengths 
has a lenticular shape, and so differs from that of the edge-on Magellanic galaxy 
NGC 4236, which has a similar luminosity and distance (Davidge 2018b). 
The disk of NGC 55 is lop-sided, with what appears to be a tidal arm 
extending in the general direction of its nearest large companion -- NGC 300. 
Isolated HI clouds are found along the disk plane at projected distances 
of $\sim 30$ kpc from the main body of the galaxy (Westmeier et al. 2013).
The HI disk of NGC 55 also shows signs of being disturbed, hinting at a chaotic 
past (Westmeier et al. 2013). 

	Despite having features that might be associated with tidal 
interactions, NGC 55 is not in a dense environment. NGC 55 and 
NGC 300 are the dominant members of a sub-system that is on the 
near side of the so-called Sculptor Group. The dwarf galaxies ESO294-010 and ESO410--005 
are other possible members of this sub-group, which has an estimated crossing time 
that exceeds the Hubble time (Karanchentsev et al. 2003). 
Other members of the so-called Sculptor Group are at larger distances from NGC 55 than 
might be inferred simply from their projected location on the sky. 

	Source confusion that is exacerbated by the orientation of NGC 55 on the sky 
makes it challenging to resolve individual stars except in the outer areas 
of the stellar disk and in the extraplanar regions. Still, NGC 55 was one of the 
first galaxies outside of the Local Group for which the detection of 
individual stars was claimed (Graham 1982). NGC 55 was also an early target of 
narrow-band photometric surveys to identify C stars (Pritchet et al. 1987). The C star 
frequency deduced from those data is significantly higher than in the LMC and SMC 
(Battinelli \& Demers 2005), suggesting levels of star-forming activity 
in NGC 55 during intermediate epochs that exceed those in the LMC and SMC. 
Davidge (2018a) examined the spectroscopic properties of the northern 
half of the NGC 55 minor axis, and found that the Ca triplet lines 
near the disk plane have depths that are consistent with 
[Fe/H] $=-0.35$ and a luminosity-weighted age of 1 -- 2 Gyr. A large C star 
frequency might then be expected. 

	Magrini, Goncalves, \& Vajgel (2017) measure 
oxygen abundances in HII regions along the NGC 55 disk plane that 
are intermediate between those of the LMC and SMC (e.g. Carlos Reyes et al. 2015). The 
analysis of the spectra of bright supergiants in NGC 55 reveals LMC-like abundances 
(Castro et al. 2012; Kudritzki et al. 2016; Patrick et al. 2017). While there appears 
not to be a radial abundance gradient in the LMC (e.g. Toribio San Cipriano et al. 
2017; Pagel et al. 1978), there is conflicting evidence about radial abundance trends 
in NGC 55. Kudritzki et al. (2016) find evidence for a metallicity gradient in the 
interstellar medium (ISM), while Magrini et al. (2017) find a flat radial abundance 
profile. Kudritzki et al. (2016) conclude that NGC 55 is accreting large amounts of gas. 
If this is the case then the chemical content of HII regions may 
not mirror that of older stars in the same part of the galaxy.

	While HII regions are found throughout NGC 55, 
much of the present-day star-forming activity is concentrated in two star-forming 
complexes in the central kpc of the galaxy that have 30 Dor-like luminosities 
(Ferguson et al. 1996). Otte \& Dettmar (1999) label these as H2 and H4 in their Figure 2. 
Recent star formation in these -- and other -- complexes has undoubtedly helped to shape 
the present-day spatial distribution of the ISM in NGC 55, including the excitation of a 
diffuse interstellar component that is found over much of the galaxy (e.g. Graham \& 
Lawrie 1982; Hoopes et al. 1996; Ferguson et al. 1996; Otte \& Dettmar 1999).

	A substantial extraplanar ISM is present in NGC 55, and 
the properties of this gas provides additional insights into the evolution of the galaxy. 
The HI distribution of NGC 55 is not symmetric about the major axis 
(Westmeier et al. 2013), and the presence of bubbles, shells, and chimney structures in 
the ISM indicate that material is being heated and expelled from the disk. 
HII regions are found well off of the disk plane (Tullmann et al. 2003; Kudritzki et al. 
2016), and Tullmann et al. (2003) measure metallicities that are 
lower than in the disk. Such low metallicities indicate that mechanisms other than the 
ejection of gas from the disk contribute to the extraplanar gas and dust. 
A population of HI clouds have been found around NGC 55 
(Westmeier et al. 2013), and these might serve as a reservoir for fueling extraplanar 
star formation. 

	Davidge (2005) used deep GMOS images to probe the stellar 
content off of the disk plane, and found that the extraplanar 
environment is dominated by old stars. While finding old stars in this part of NGC 55 
is not unexpected, the color of the extraplanar RGB is consistent with [M/H] $\sim -1.2$ 
to $-1.3$. This is similar to the metallicity found in extraplanar HII regions, hinting 
at a flat age-metallicty relation in this part of the galaxy. Tikhonov et al. (2005) 
estimate a similar metallicity from deep HST images of the same area. 

	It is apparent that NGC 55 has characteristics 
that may not be typical of Magellanic Irregular galaxies.
In the present paper we discuss a grid of long-slit spectra 
of NGC 55 recorded with the Gemini Multi-Object Spectrograph (GMOS) on Gemini South (GS). 
A contiguous $0.4 \times 5$ arcmin area was observed, centered near the intersection of the 
major and minor axes. The target area covers regions to the south east of the minor axis 
that have been the subject of many previous studies (see above), as well as the 
less-studied area to the north west of the minor axis. 

	The primary goal is to examine the luminosity-weighted 
properties of the central regions of the galaxy, which is an environment that 
is expected to be rich in information about its past evolution. 
Integral field observations of this nature are a powerful means of
investigating stellar content and galactic structure in complicated environments 
(e.g. Roth et al. 2018). The scientific benefits of integral field 
observations has motivated the development of instruments such as 
MUSE (e.g. Bacon et al. 2010), as well as software packages such as TYPHOON (Sturch \& 
Madore 2012).

	Spectra of compact sources with a FWHM that is comparable to that of the seeing 
disk, and so are unresolved, are also examined. With an image scale 
of 1 arcsec $\sim 10$ parsecs, then the majority of these sources are 
almost certainly blended asterisms, star clusters, or compact HII regions. 
The GMOS spectra are supplemented with archival GALEX (Martin et al.  2005), 2MASS 
(Skrutskie et al. 2006), and SPITZER (Werner et al. 2004) images. The GALEX images trace 
star formation over the past few tenths of a Gyr, while the 2MASS and SPITZER images trace 
stars that formed over a much longer time span and are the dominant contributor 
to overall stellar mass. 

	Details of the observations and the steps used to remove instrumental and 
atmospheric signatures are reviewed in Section 2. The GALEX, 2MASS, and Spitzer 
observations are discussed in Section 3, while the projected distribution of 
emission and absorption lines in the area surveyed with GMOS, including the spectroscopic 
properties of the largest star-forming complexes, are the subject of 
Section 4. The nature of unresolved sources are examined in Section 
5. The paper closes in Section 6 with a discussion and summary of the results. 

\section{OBSERVATIONS \& REDUCTIONS}

\subsection{Description of the Observations}

\begin{table*}
\begin{center}
\caption{Dates of Observation}
\begin{tabular}{ll}
\startdata
\tableline\tableline
Pointing & Date Observed \\
 & (UT 2017) \\
\tableline
p10 & July 13 \\
p8 & July 11 \\
p6 & July 10 \\
p4 & July 10 \\
p2 & July 3 \\
Center & July 3 \\
n2 & July 13 \\
n4 & July 13 \\
n6 & July 13 \\
n8 & July 13 \\
n10 & July 24 \\
n12 & July 24 \\
 & \\
Minor Axis & July 3 $+$ July 24 \\
\tableline
\end{tabular}
\end{center}
\end{table*}

	The spectra discussed in this paper were recorded with GMOS 
(Hook et al. 2004) on GS for program GS-2017A-Q-98 (PI: Davidge). 
The GMOS detector is a mosaic of three $2048 \times 4176$ Hamamatsu CCDs. With on-sky 
sampling of 0.080 arcsec pixel$^{-1}$, the rectangular detector mosaic allows spectra 
to be recorded at wavelengths that are dispersed out of the $5.5 \times 5.5$ arcmin 
sky footprint. The detector was binned $4 \times 4$ pixels during read-out 
given the image quality and spectral resolution.

	Individual spectra were recorded through a 5.5 arcmin long and 2 arcsec wide slit. 
The light was dispersed with the R400 grating (400 l/mm; $\lambda_{blaze} = 7640\AA$, 
$\lambda/\Delta \lambda \sim 500$ with a 2 arcsec slit). The grating was 
rotated so that spectra were recorded with wavelengths at the detector center of either 
8200 or 8500\AA . Recording spectra with two central wavelengths allows holes in 
wavelength coverage due to gaps between CCDs to be filled during processing. 
An OG515 filter was deployed to suppress light from higher orders.

	Thirteen contiguous pointings were observed that map the 
brightest central parts of NGC 55. Twelve of the pointings run parallel 
to the major axis, and these are refered to 
as the `disk' pointings. The disk pointings sample the galaxy 
in 2 arcsec increments perpendicular to the major axis of NGC 55. The 
disk pointings together cover a $0.4 \times 5$ arcmin$^2$ area. The area observed is shown 
in Figure 1 against the backdrop of a SPITZER [3.6] image. The central co-ordinates and the 
dates of observation for each pointing are listed in Table 1. The naming convention in 
this table and for the remainder of the paper is such that `p' and `n' refer to 
positive and negative offsets perpendicular to the major axis. 
The number following 'p' or 'n' is the offset in arcsec from the 
major axis. The Center field runs along the major 
axis of the galaxy. Four 450 sec exposures were recorded at each disk pointing, 
with two exposures at each central wavelength.

\begin{figure*}
\figurenum{1}
\epsscale{0.7}
\plotone{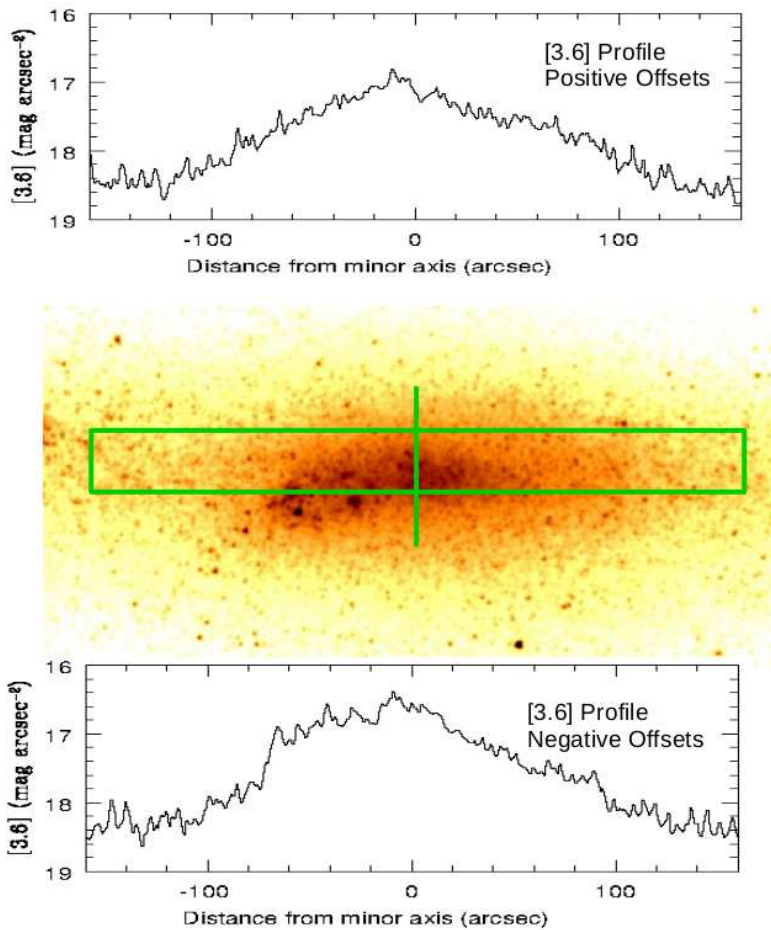}
\caption{(Middle panel) [3.6] image of NGC 55, showing the area observed with GMOS. 
The image was recorded for the program described by Sheth et al. (2010), and was 
downloaded in processed form from the NASA Extragalactic Database. The image 
has been rotated so that the NGC 55 disk plane lies along 
the horizontal axis. North is towards the upper right hand corner, 
and East is towards the upper left hand corner. The green box 
has a length of 5 arcmin and a width of 0.4 arcmin, and 
shows the area covered by the disk pointings. The vertical green line marks the 
slit location for the minor axis pointing. (Upper and lower panels) Mean [3.6] light 
profiles of the regions that have positive and negative 
offsets from the major axis. The differences between the 
two light profiles foreshadows the diversity of properties in the spectra. 
The light profile defined from positive offsets -- where there is little or no 
contamination from the star-forming complexes H2 and H4 -- follows an exponential 
on each side of the minor axis. The negative offset light profile is less well-defined 
and is not symmetric about the minor axis; aside from a bump to the left 
of the minor axis that is due to H2 and H4, the surface brightness near the 
the minor axis is a few tenths of a magnitude arcsec$^{-2}$ higher than that 
extrapolated from the disk light profile at larger positive radii.} 
\end{figure*}

	Observations were also recorded with the slit placed along the minor axis. 
NGC 55 is viewed almost edge-on, and has a moderately 
compact light distribution perpendicular to the disk plane. Minor axis 
spectra were recorded with the galaxy stepped between 
two locations on the slit, so that the sky could be removed by subtracting 
succesive spectra recorded at different step locations. Minor axis spectra were 
recorded on two nights, and eight 450 sec exposures were recorded in total. 
Negligible sky residuals in the processed minor axis spectrum 
at wavelengths where strong emission lines dominate 
the night sky signal are testament to the stability of 
the sky level when these data were recorded. 
The minor axis spectrum was discussed previously by Davidge (2018a).

	The disk pointing spectra were not stepped along the slit, as the 
major axis of NGC 55 has a length that exceeds the GMOS slit length. While background sky 
is not sampled, the sky level for each pointing can 
still be estimated by comparing the signal from the sky-subtracted 
minor axis spectrum with the sky$+$galaxy spectrum at 
the point where the minor axis and disk spectra intersect (Section 2.3). An alternate 
observing strategy would be to apply large angular offsets immediately following individual 
disk observations and record spectra of pristine sky. However, this lowers efficiency, as 
only half of the exposures would then sample the science target. Additional time is 
also required to offset and re-acquire the science field. 

	Calibration exposures that are required to remove instrumental and 
atmospheric signatures from the spectra were also recorded. Calibrations that 
did not require time on-sky include: (1) bias frames that are recorded 
on a daily basis by Gemini staff; (2) images of dispersed light from a continuum 
source in the facility calibration unit (GCAL) that were recorded midway through 
observing sequences; and (3) spectra of a CuAr arc that is also in GCAL, recorded at 
the end and/or beginning of nights. The arcs were recorded 
with the same central wavelengths as the science 
spectra. On-sky calibrations consisted of (1) spectra of the dO star Feige 110 to 
gauge contamination from telluric absorption features and wavelength response, 
and (2) dispersed light of the twilight sky that are used to construct a 'twilight' flat 
field frame. The twilight flat is used to correct for differences in slit 
illumination between sources on the sky and the flat-field lamp in GCAL (Section 2.2).

	Neither the NGC 55 disk nor Feige 110 spectra were recorded 
during photometric conditions. Hence, absolute fluxes can not be determined from 
the NGC 55 spectra. Still, as the NGC 55 spectra are contiguous and all intersect 
the minor axis spectrum then they can be calibrated in a self-consistent internal 
manner, such that the shape (but not the zeropoint) of the spectral energy distribution 
is retained.

\subsection{Initial Processing: Removal of Instrumental Signatures}

	All processing was done using routines in the 
IRAF \footnote[1]{IRAF is distributed by the National Optical Astronomy Observatories,
which are operated by the Association of Universities for Research
in Astronomy, Inc., under cooperative agreement with the National
Science Foundation.} data reduction package. The initial processing removed static 
instrument signatures that originate in the optics and detector. 
To start, a median bias frame was constructed for each night that data 
were recorded, and the result was subtracted from all exposures obtained 
on that night. The GMOS CCDs are read out with twelve amplifiers, 
and the bias-subtracted data were adjusted for amplifier-to-amplifier 
differences in gain. Bad pixels and columns were identified using 
existing maps of these devices. Cosmic-rays were found by 
identfying pixels that fell above the median value of surrounding pixels by a pre-defined 
threshold. The threshold was selected to prevent triggering on noise events and 
bright emission features in the science spectra. Bad pixels and cosmic rays were 
repaired by interpolating between `good' pixels.

	Flat-fielding was done in two steps. First, each spectrum was divided by 
flat-field frames that used the GCAL continuum lamp as the source of 
illumination. These flat-field frames have a high S/N ratio, and correct 
for pixel-to-pixel and CCD-to-CCD differences in quantum efficiency as well as 
for vignetting along the slit. However, the optical feed from GCAL does not follow the same 
light path as signal from the sky, and so the slit is illuminated 
by GCAL in a slightly different manner than the science data. This is remedied in the 
second flat-fielding step, in which the twilight sky flat is used to compensate for 
differences between GCAL and sky illumination. Absorption features in the twilight flat, 
which is a dispersed spectrum of scattered sunlight, were suppressed by collapsing the 
twilight flat along the dispersion axis after division 
by the GCAL flat. This wavelength smoothing also increased the signal-to-noise 
ratio of the twilight flat. The result is an achromatic illumination correction 
that was applied to the science data.

	The final step in the initial processing was to correct for 
geometric distortions introduced by the GMOS optics. These cause 
emission and absorption features that should be straight in the direction 
perpendicular to the dispersion axis to instead appear curved. 
A distortion map was constructed by tracing bright emission lines in 
a mean arc spectrum, and the result was used to define a 
two-dimensional warping function to correct for these distortions. This relation 
was then applied to the science data.

\subsection{Final Processing: Construction of Spectra for Analysis}

	The goals of the second phase of the processing are to remove telluric 
emission and absorption features, and adjust for instrumental 
variations in throughput. The spectra are also wavelength calibrated. 
The end product are spectra that can be used to examine the spectroscopic characteristics 
of emission and absorption features throughout the central regions of NGC 55.

	The subtraction of the sky from the minor axis data was 
achieved by differencing pairs of observations recorded with the same central 
wavelength, but at different locations along the slit. Since the sky level 
is measured at the same location on the slit as the 
science data (albeit at slightly different times) then 
systematics that arise due to detector fringing and flat-fielding are suppressed.
The resulting differenced two-dimensional spectra 
contain positive and negative versions of the target spectrum. 
Common spatial intervals were extracted from 
the negative and positive components, and a co-added spectrum was produced 
for each wavelength setting by subtracting the negative spectrum 
from the positive spectrum. Each co-added spectrum was 
wavelength calibrated, and the results at each wavelength setting were 
averaged together. Holes in the wavelength coverage due to gaps between CCDs were 
filled using signal from the other wavelength setting. 

	Sky emission in the disk spectra was removed using 
the sky-subtracted minor axis spectrum as a reference. 
After adjusting for variations in sky transparency, the 
difference between the sky-subtracted minor axis spectrum and the disk 
spectrum at the point of intersection is the sky spectrum for that disk pointing. 
Sky subtraction was checked by generating a light profile from the 
sky-subtracted spectra near $0.85\mu$m, and comparing the results to that obtained 
from 2MASS J-band images (Section 3). Good agreement between the light profiles was 
found. The sky-subtracted disk spectra were then wavelength 
calibrated and combined, with wavelength intervals that lacked 
signal again filled with spectra taken at the other wavelength setting.

	The NGC 55 spectra were divided by the spectrum of Feige 110, which 
had been processed in parallel with the NGC 55 data. Dividing by the Feige 110 spectrum 
suppresses telluric absorption features and wavelength-dependent 
throughput variations. The wavelength response of 
the NGC 55 spectra is also normalized to that of a dO star, thereby simplifying 
the identication of the continuum. H$\alpha$ absorption is present in the 
Feige 110 spectrum, and this was removed by fitting this part of the spectrum with a 
Vogt profile and then subtracting the fit. Lines of the Paschen series 
are not evident in the Feige 110 spectrum.

	Two sets of spectra were constructed for NGC 55. For one set a low-order 
continuum function was fit to each ratioed NGC 55 spectrum, and the spectra were divided by 
the results. These continuum-normalized spectra are used to measure the equivalent widths 
of various lines, and are the basis for spectra displayed in figures throughout the paper. 
The generate the second set, the ratioed spectra were multiplied by a response function 
that recovered the spectral energy distribution (SED) of each offset spectrum, using the 
Feige 110 spectrum as a reference. These spectra are used to determine relative line 
strengths.

\section{GALEX, 2MASS, and SPITZER OBSERVATIONS}

	Archival ground and space-based images that span a broad range of wavelengths 
provide supplemental information for understanding the evolution of NGC 55. 
Images of NGC 55 taken as part of surveys and science programs conducted 
with GALEX and SPITZER, as well as images recorded as part of the 
2MASS survey, are considered here. The angular resolutions of the GALEX and SPITZER 
datasets are comparable to those of the GMOS spectra. 

	The SPITZER observations are [3.6] and [4.5] images discussed by Sheth et al. 
(2010). With the exception of luminous red supergiants (RSGs) and AGB 
stars that belong to young and moderately young populations, much of the light in 
[3.6] and [4.5] originates from older giant branch stars that trace 
stellar mass. Dust extinction is a concern for NGC 55 as it 
is a star-forming galaxy that is viewed almost edge-on, and 
absorption by dust along the line-of-sight is greatly reduced 
in [3.6] and [4.5] when compared with images in the visible and UV. 

	$J$ and $Ks$ images of NGC 55 were taken from the 2MASS Large Galaxy 
Atlas (Jarrett et al. 2003). A significant contribution to the integrated light in 
these filters comes from older evolved red stars that trace the 
stellar mass of the galaxy. The 2MASS images have $\sim 3$ arcsec 
FWHM angular resolution with a shallow photometric depth, and so their 
use in the current study is restricted to broad-band color measurements near the center of 
the galaxy. Finally, FUV and NUV images from the GALEX 
Ultraviolet Atlas of Nearby Galaxies (Gil de Paz et al. 2007) are also examined. These 
highlight areas of recent star formation as well as pockets of high dust obscuration. 

	Processed images from all three datasets were downloaded from the NASA 
Extragalactic Database \footnote[2]{http:www.ned.ipac.caltech.edu}. These 
were rotated to place the major axis of NGC 55 along the horizontal axis, and then 
re-sampled to match the pixel scale of the SPITZER data. Sky levels were 
measured in areas that are free of light from NGC 55, and the results were 
subtracted from the images.

	There are areas of extended emission in the [3.6] image 
in Figure 1, some of which correspond to concentrations of 
hot stars that are also visible in the GALEX images (see below). Individual unresolved 
luminous sources are also sprinkled across the [3.6] image. The intrinsic spatial 
resolution of the SPITZER data is coarse ($\sim 20$ pc at the distance of NGC 55, based on 
2 arcsec FWHM image quality measured in the [3.6] images), and so any apparent point 
sources likely contain signal from more than one star. This being said, highly evolved 
stars are intrinsically luminous objects with comparatively low effective temperatures, 
and so stand out in [3.6] and [4.5] with respect to the background light that 
originates predominantly from warmer objects. The contrast with 
respect to background light is further enhanced 
if the evolved stars are surrounded by hot circumstellar dust, as can occur around 
intermediate mass stars that are nearing the end of their evolution. 
In Section 5 it is shown that some unresolved sources in the Spitzer observations 
have spectroscopic signatures that are indicative of highly evolved AGB stars.

	Light profiles of the positive and negative disk pointings were constructed  
from the [3.6] observations by summing the light in each angular interval on 
either side of the major axis, and the results bracket the [3.6] image 
in Figure 1. In both cases the profile to the right of the minor axis 
defines a disk-like exponential, suggesting that single sources do not dominate the 
light profile in this part of the galaxy. However, the positive and negative profiles 
to the left of the minor axis are different. The areas of star formation 
in and around the sources H2 and H4 (Otte \& Dettmar 1999) produce prominent 
peaks in the left hand side of the negative offset light profile. The light from H2 and H4 
in [3.6] likely originates from a mix of luminous red 
stars, thermal emission, and -- if there are stars present with 
ages $\leq 4$ Myr -- nebular continuum emission (e.g. Figure 12 of Byler et al. 2017).

	There is also a modest structure in the negative 
offset profile that is centered on the minor axis. 
An extrapolation of the portion of the negative profile that is to the right of the 
minor axis to the center of the galaxy indicates that the area within 
$\pm 15 - 20$ arcsec of the minor axis has a surface brightness that is 
a few tenths of a magnitude arcsec$^{-2}$ higher than what is expected for a purely 
exponential disk. There are pockets of relatively deep Ca triplet 
absorption in this part of the galaxy that are accompanied by 
luminous red stars in the [3.6] image (Section 4).

	In contrast to the negative offset profile, the positive offset profile 
follows an exponential to the left of the minor axis, but 
flattens near the left hand edge of the panel due to 
light from the tail of stars and gas that emerges from this part of 
the galaxy. Similar behaviour at large radii to the left of the minor axis is 
seen at negative offsets. The distinct break in the light profiles indicates 
that the tail of stars and gas is structurally distinct 
from the main body of the disk, as might be expected if it is a tidal feature.

	The diverse range of environments and populations sampled by GMOS becomes 
apparent when examining the broad-band colors generated from the images. 
FUV--NUV, J-K, and [3.6]--[4.5] were computed with the native angular 
resolution of each dataset (i.e. not balanced over the entire suite of images), 
and the results are shown in Figure 2. Areas with the bluest colors are shown in 
white, while darker shades correspond to progressively redder colors.

\begin{figure*}
\figurenum{2}
\epsscale{1.0}
\plotone{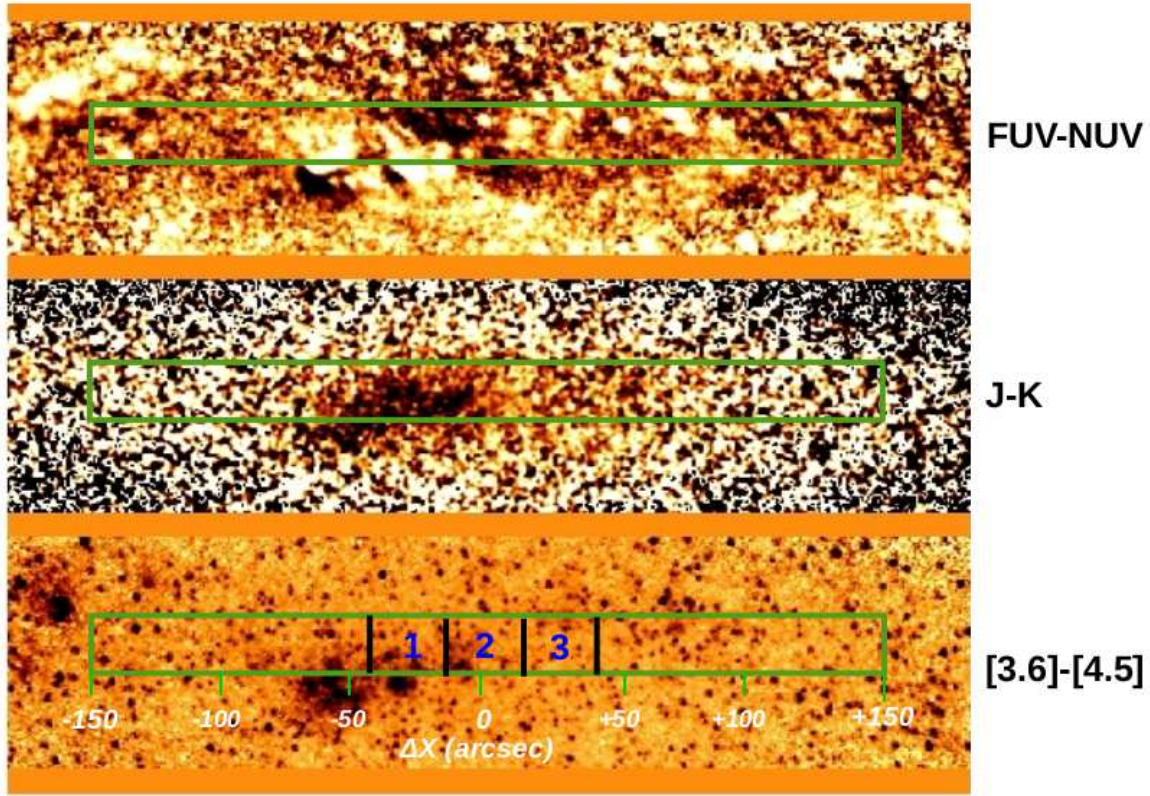}
\caption{FUV--NUV (top panel), J--K (middle panel), and [3.6]--[4.5] (bottom panel) 
colors near the center of NGC 55. The green box in each panel marks the area 
observed with GMOS. Areas with the bluest colors are shown in white, while the 
darkest areas have the reddest colors. The three angular intervals used to 
construct the spectra shown in Figure 3 are marked in 
the bottom panel, along with offsets from the minor axis 
($\Delta$X). The complex distribution of the light near the 
center of NGC 55 is evident in the broad-band colors. Concentrated pockets of 
recent star-forming activity with blue FUV--NUV colors due to early-type 
stars, and red colors in $J-K$ and [3.6]--[4.5] due to thermal and continuum 
emission, are obvious to the left of the minor axis. The area to the 
right of the minor axis samples a part of the galaxy that has 
been less active in terms of recent star formation, although pockets of hot blue 
stars are still present, and these are seen in the GMOS spectra (Section 4).}
\end{figure*}

	The mottled appearance of the FUV--NUV map in Figure 2 highlights 
the spatial distribution of hot, young stars and star-forming 
regions, coupled with non-uniformities in line-of-sight extinction. 
FUV--NUV colors indicate that areas of large-scale star formation occupy much of the 
area to the immediate left of the minor axis in the GMOS field, while 
smaller, more isolated star-forming regions are present to the right of the minor axis. The 
large-scale serpentine-like structure in the left hand half of the FUV-NUV map 
appears to track bubbles and areas of star formation triggered by outflowing material 
that has been identified in previous studies. 

	There is less variation in $J-K$ in the area to the right of the minor axis in 
Figure 2 than to the left of this axis. There is a gradient in $J-K$ moving 
away from the large star-forming complexes along both the major and 
minor axes, in the sense of J--K becoming smaller (lighter colors in Figure 2) with 
increasing distance. While there is nebular emission combined with 
non-uniform obscuration near the center of NGC 55, the overall trend in $J-K$ 
likely is due to systematic variations in stellar content. If such variations 
in stellar content are present then a gradient in the depth of the Ca triplet might 
be expected. In fact, Davidge (2018a) finds a gradient in the depth of the Ca 
triplet along the minor axis of NGC 55. This gradient was attributed 
to metallicity decreasing with increasing distance from the major axis, 
and the sense of the $J-K$ gradient is consistent with this. 
In summary, the $J-K$ and FUV--NUV color distributions 
in Figure 2 indicate that the areas to the right of the minor 
axis track more closely the disk mass profile of NGC 55 than those to the left of the 
minor axis, as they are less affected by areas of recent lare-scale star formation.

	When considered over angular scales of a few tens of arcsec or more, the 
[3.6]--[4.5] color is uniform across much of NGC 55, due to 
the muted sensitivity of the Rayleigh-Jeans tail of 
the SEDs to temperature variations, coupled with the lower sensitivity of integrated colors 
at these wavelengths to reddening. Still, there are spatially extended areas that have 
very red [3.6]--[4.5] colors that coincide with star-forming regions in the FUV--NUV map. 
The red [3.6]--[4.5] color in these regions is likely due to dust that has been 
heated by young, hot stars. There are also a number of discrete sources 
scattered across the [3.6]--[4.5] map. These are the same 
unresolved sources discussed earlier in this section, and these 
are more conspicuous in Figure 2 because the large-scale light variation due to the main 
body of the galaxy is suppressed when colors are displayed. 
The detection of these objects in the [3.6]--[4.5] map is consistent 
with them being highly evolved stars or clusters in which the light is dominated by 
luminous red stars.

\section{THE PROJECTED DISTRIBUTION OF EMISSION AND ABSORPTION FEATURES}

	The images discussed in Section 3 indicate that the 
area examined with GMOS samples a broad mix of populations and environments. 
This is further demonstrated in Figure 3, where the mean spectra of 
three angular intervals along the major axis are compared. The areas over 
which the mean spectra were generated are marked in the bottom panel 
of Figure 2.

\begin{figure*}
\figurenum{3}
\epsscale{1.0}
\plotone{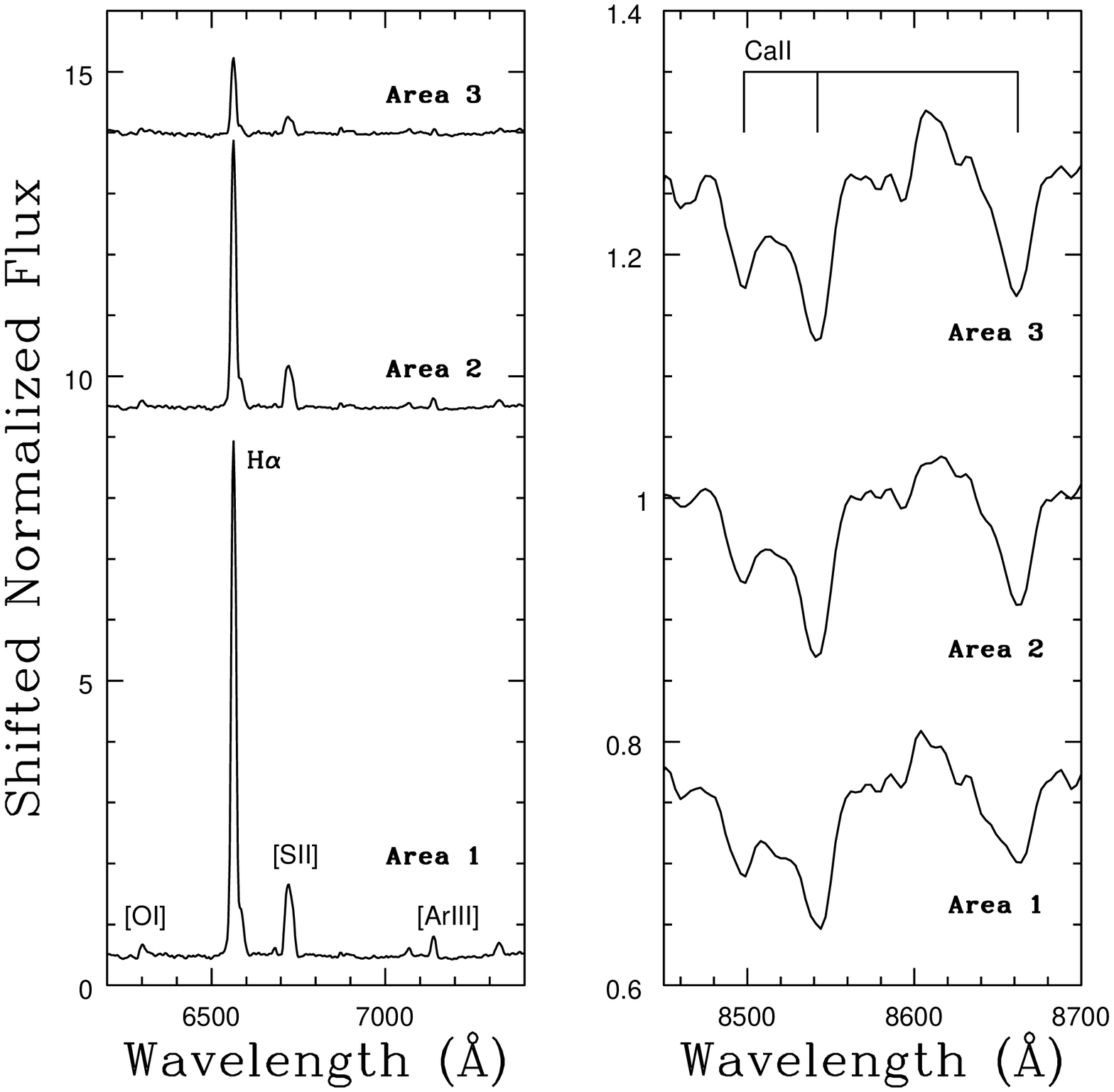}
\caption{Mean spectra of the three angular intervals centered on the minor axis of NGC 55 
that are marked in the lower panel of Figure 2. Wavelengths near H$\alpha$ and the Ca 
triplet are shown. The spectra have been normalized to the continuum, and shifted 
along the vertical axis for display purposes. There is a gradient in emission line 
strengths from Area 1 to Area 3. The strong emission lines in the 
Area 1 spectrum originate in the H4 star-forming complex. That the variations in the 
Ca lines are much more subtle than those among the emission lines suggests that 
veiling by continuum emission near 8600\AA\ is modest.}
\end{figure*}

	Area-to-area differences in emission line strengths 
are evident in the left hand panel of Figure 3, as might be expected from the 
color measurements discussed in Section 3. Emission lines are strongest 
in the spectrum of Area 1, which contains the H4 
star-forming complex. The FUV--NUV and [3.6]--[4.5] 
color maps indicate that there is strong UV emission combined with significant 
amounts of dust emission in Area 1. Smaller -- but nonetheless significant 
-- area-to-area differences are also seen in the depths of the Ca triplet 
lines in the right hand panel of Figure 3. That the Ca triplet lines do not 
differ markedly in these spectra suggests that veiling by 
continuum emission may not be a major factor when interpreting NGC 55 spectra near 
8600\AA .

	To trace spectroscopic characteristics, 
line indices were defined for key emission (H$\alpha$+NII; He I 6678; 
[SII]6717+6731; He I 7065; and [SIII]9069) and absorption (Ca3 and CN7900) features. 
These features were selected based on their 
strengths and astrophysical significance. The index passbands 
are listed in Table 2, where wavelength intervals that sample the target line 
(or lines) as well as continuum measurements that straddle the line(s) are specified. 
All indices are measured as equivalent widths. The 
2 arcsec slit width largely defines the angular resolution of these data, and so 
spectra were binned along the spatial direction prior to measuring 
indices so that each spaxal samples $2 \times 2$ arcsec. 

\begin{deluxetable*}{lccc}
\tablecaption{Index Passbands}
\startdata
\tableline\tableline
Index & Feature & Continuum 1 & Continuum 2 \\
 & (\AA) & (\AA) & (\AA) \\
\tableline
H$\alpha$ & 6530--6610 & 6450--6530 & 6650--6690 \\\relax
[SII]6717$+$6731 & 6702--6749 & 6689--6702 & 6749--6762 \\
HeI 7065 & 7044--7083 & 7005--7044 & 7083--7121 \\\relax
[SIII]9069 & 9040--9090 & 8938--9040 & 9090--9192 \\
 & & & \\
CN7900 & 7875--7975 & 7800--7875 & 7975--8050 \\
Ca3 & 8480--8680 & 8450--8480 & 8680--8710 \\
\tableline
\enddata
\end{deluxetable*}

\subsection{Emission Lines}

	The spatial distribution of H$\alpha$ emission in the central regions of 
NGC 55 is examined in the top panels of Figures 4 - 6. The 
numbers in the H$\alpha$ panels are line strengths, normalized to 
that in H2. The line strengths were normalized to H2 because an absolute 
flux calibration was not possible (Section 2), and this is where H$\alpha$ is strongest. 
The line strengths were measured over $6 \times 6$ arcsec$^2$ 
areas to boost the signal to noise ratio near the slit edges, where the signal 
is lowest. The H$\alpha$ measurements are displayed with a logarithmic stretch.

	The relative strengths of other emission features, normalized to H$\alpha$, 
are also shown in Figures 4 - 6. The line ratios are displayed with a linear stretch, and
the numbers shown in these panels reflect signal averaged over $6 \times 6$ 
arcsec$^2$ areas. There is a tendency for the H$\alpha$ measurements and line ratios 
near the right hand edge of Figure 6 to depart markedly from 
values that are more typical for the region observed with GMOS, 
and this is due to the low surface brightness near the slit edges (Figure 1).

\begin{figure*}
\figurenum{4}
\epsscale{1.0}
\plotone{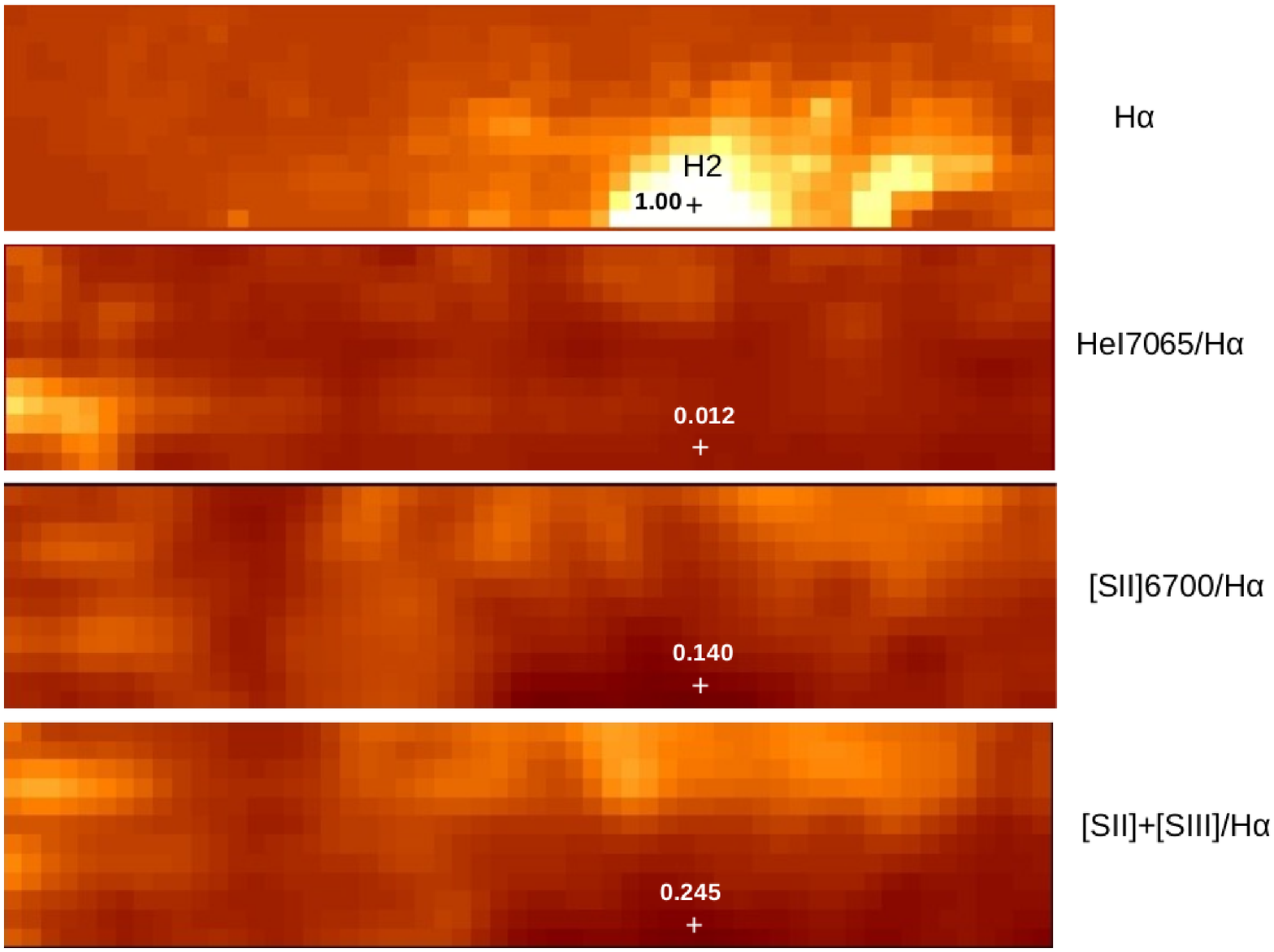}
\caption{Spatial distribution of H$\alpha$ line strengths (top 
panel), and the relative strengths of three emission lines with respect to 
H$\alpha$ (lower three panels). The angular interval examined in this figure 
is the left hand third of the GMOS field shown in Figure 1. 
The numbers in the top panel show the relative strength 
of H$\alpha$ emission normalised to that of H2. The numbers 
in the other panels show relative line strengths with respect to H$\alpha$.
[SII]$+$[SIII] is the sum of the [SII] 6717$+$6731 \AA\ doublet and 
the [SIII]9069 line strengths. H$\alpha$ measurements 
are displayed with a logarithmic stretch while line ratios are shown with 
a linear stretch. The star-forming complex identified as H2 
in Figure 2 of Otte \& Dettmar (1999) is marked.} 
\end{figure*}

\begin{figure*}
\figurenum{5}
\epsscale{1.0}
\plotone{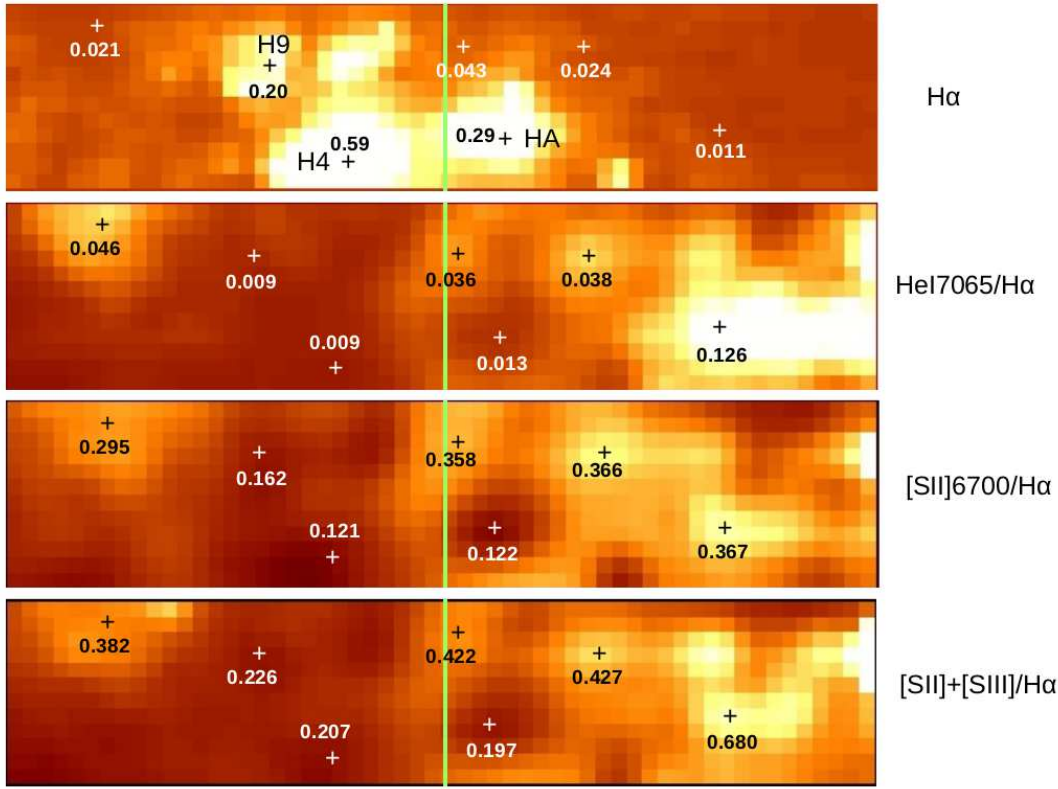}
\caption{Same as Figure 4, but for the middle third of the area observed with GMOS. 
The surface brightness at visible/red wavelengths is highest in this area. 
The minor axis of NGC 55 is indicated by the green 
line. The star-forming complex H4 (Figure 2 of Otte \& Dettmar 
1999) dominates the H$\alpha$ distribution, although smaller H$\alpha$ 
emitting areas are also present. There is a tendency for HeI7065/H$\alpha$, 
[SII]6717$+$6731/H$\alpha$, and [SII]$+$[SIII]/H$\alpha$ to peak outside of the 
star-forming regions. The prominent ring-shaped structure in the 
HeI/H$\alpha$, [SII]/H$\alpha$, and [SII]$+$[SIII]/H$\alpha$ diagrams 
immediately to the right of the minor axis and centered on HA is discussed in the 
text.}
\end{figure*}

\begin{figure*}
\figurenum{6}
\epsscale{1.0}
\plotone{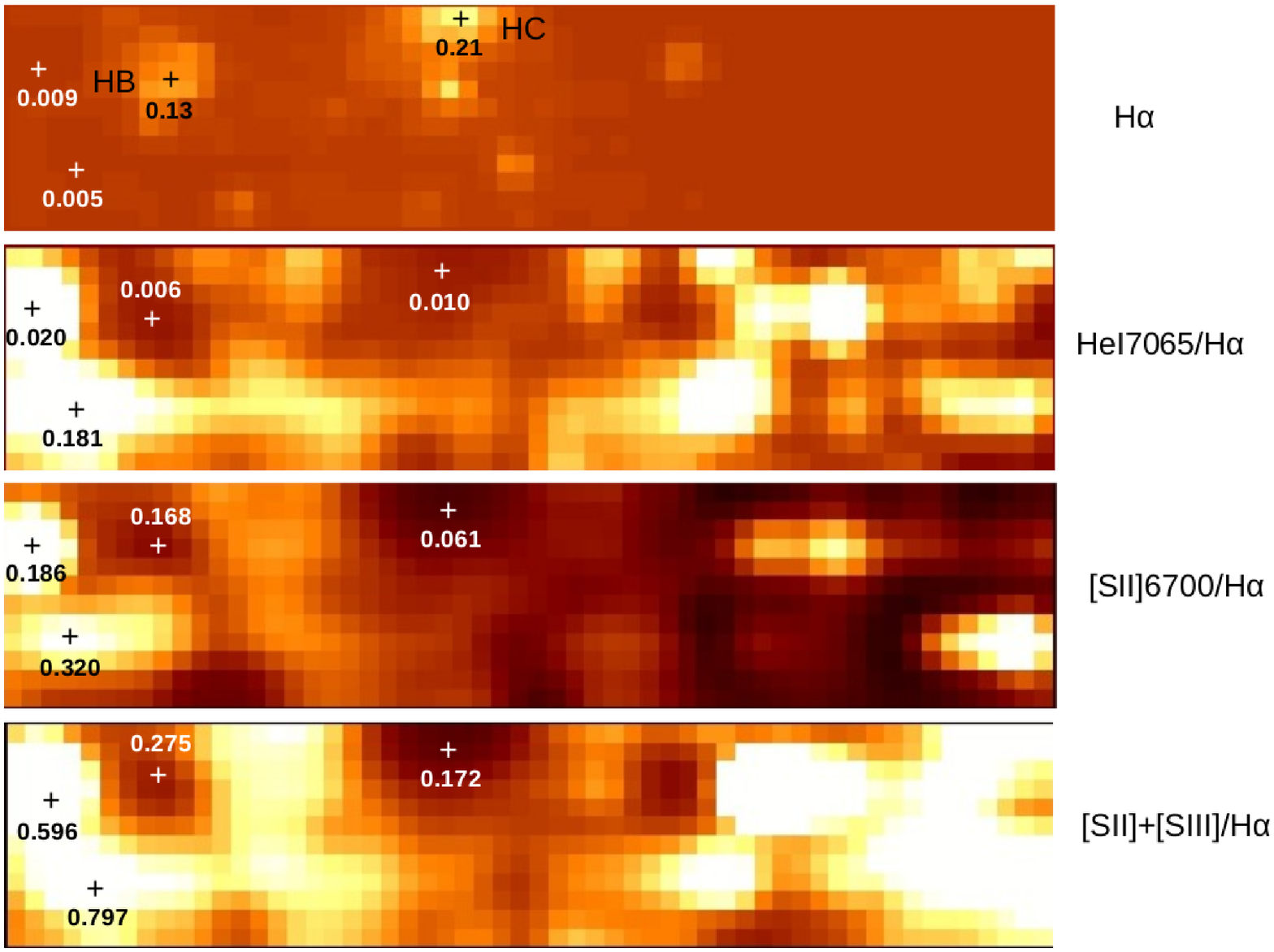}
\caption{Same as Figure 4, but for the right hand 
third of the area observed with GMOS in Figure 1. Localized areas of star 
formation dominate the H$\alpha$ map, and HB and HC are star-forming regions that are 
discussed in the text. The surface brightness in this part of NGC 55 at visible/red 
wavelengths is lower on average than in the areas displayed in Figures 4 and 5.
Line measurements that depart markedly from those that are more 
typical throughout the area observed with GMOS 
tend to occur where low signal levels affect line strength measurements.}
\end{figure*}

	H$\alpha$ is blended with [NII]6584 in these 
spectra. The contribution that [NII] makes to the combined H$\alpha$+[NII] signal 
was estimated using the IRAF task SPLOT. The H$\alpha$ and [NII]6584 lines were 
both fit with two Gaussians having the same width, and the results indicate that [NII] 
contributes $\sim 6\%$ of the total signal in the central regions of 
NGC 55, with a $\pm 1\%$ dispersion. This is 
consistent with line strengths measured in NGC 55 HII regions by Magrini 
et al. (2017), where [NII] contributes $\sim 5\%$ on average to 
the combined H$\alpha$ and [NII]6584 flux, with a standard deviation of $\pm 3\%$. 
The measurements made by Webster \& Smith (1983) have a mean [NII]/H$\alpha = 0.06$. 
Therefore, while [NII]6584 contributes to the H$\alpha$ 
measurements in Figures 4 -- 6, it is at only at the few percent level. 

	Figure 1 of Graham \& Lawrie (1982) shows the distribution of [OIII]5007 emission 
in part of the area observed with GMOS, and the distribution of [OIII] 
emission is similar to that defined by the H$\alpha$ index 
in Figures 4 and 5\footnote[3]{Much of the area covered in Figure 6 falls 
outside of the area observed by Graham \& Lawrie (1982).}.
A feature that Graham \& Lawrie identify as the 'outer arc' 
originates near the left hand edge of Figure 5, and is resolved 
into sub-structures in Figure 2 of Otte \& Dettmar (1999). 
While the outer arc is not obvious in the H$\alpha$ measurements, pockets 
of bright HeI7065 and [SII] emission in Figure 5 coincide with this feature.

	The HeI 7065 line is one of the strongest He emission lines in the 
red part of the spectrum, and the HeI7065/H$\alpha$ ratio 
gauges variations in ionization conditions. Significant variations 
in the HeI7065/H$\alpha$ ratio are seen in Figures 5 and 6.
The HeI7065/H$\alpha$ ratio tends to be lowest in HII regions, and there is only 
modest nebula-to-nebula variation in this ratio. The one exception is HC, where 
HeI7065/H$\alpha$ is lower than in other star-forming regions, suggesting that ionization 
conditions in HC differ from those in the other nebulae.

	The complex nature of the ISM in the central regions of NGC 55 is evident in the 
[SII]/H$\alpha$ ratios, and in Figure 5 it is clear that the strengths of [SII] 
and H$\alpha$ are anti-correlated. This anti-correlation is a well known 
phenomenon in diffuse interstellar gas (DIG), and was first investigated in NGC 55 by 
Hoopes et al (1996). If the DIG is excited by photoionization from Lyman photons 
then the [SII]/H$\alpha$ ratio should increase with distance from the ionizing stars. 
Ferguson et al. (1996) find (1) a relation between [SII]/H$\alpha$ and 
H$\alpha$ surface brightness, in the sense of decreasing [SII]/H$\alpha$ with 
increasing H$\alpha$ surface brightness, and (2) that 
the [SII]/H$\alpha$ ratio in the main body of NGC 55 is $\sim 2\times$ higher than in 
individual HII regions. The [SII]/H$\alpha$ ratios obtained from the GMOS 
spectra show variations in excess of a factor of 2, with localized areas where the 
[SII]/H$\alpha$ ratio approaches 3 $\times$ that in HII regions. 

	A high [SII]/H$\alpha$ ratio is a signature of shocked gas, and 
[SII]/H$\alpha$ is commonly used to identify supernova remnants (SNRs). 
The [SII]/H$\alpha$ ratio in SNRs is typically $\geq 0.5$ (e.g. Blair et al. 1981). 
The vast majority of SNRs in nearby galaxies have diameters in excess of 20 pc 
(e.g. Badenes et al. 2010; Long et al. 2010), and so would appear as extended sources in 
the GMOS spectra of NGC 55. Despite the high SFR, there are no areas in 
Figures 4 -- 6 where the [SII]/H$\alpha$ threshold is breeched, although we caution 
that the numbers shown in Figures 4 -- 6 are made over angular scales that 
correspond to $\sim 60 \times 60$ parsecs.

	There is a large ring-like structure with a diameter of 200 -- 300 parsecs that is 
immediately to the right of the minor axis in Figure 5 that 
has moderately high [SII]/H$\alpha$ ratios. This structure is also 
seen in the HeI/H$\alpha$ ratio, and is more-or-less symmetric about the H$\alpha$ 
source HA, the spectrum of which is examined in Section 4.2. 
We suggest that the emission from the ring 
is powered by an outflow from the HA star-forming region.
The symmetric nature of the ring-like structure is perhaps surprising 
given its proximity to H4, as any outflow from the larger star-forming complex 
might be expected to compress the ring. Still, the orientation of NGC 55 
is such that the physical distance between the ring and H4 might be 
larger than the projected value. 

	The [SIII]9069 line is a prominent feature 
at wavelengths where the extinction from dust is 
lower than at visible/red wavelengths, although it is in a part of the spectrum that 
coincides with deep telluric water bands. The [SIII] 9069 line is significant from an 
astrophysical perspective since -- when combined with 
the strengths of the [SII] lines at 6717 and 6731 \AA\ -- 
it is an estimator of [S/H]. Vilchez \& Esteban (1996) define the S$_{23}$ 
statistic as the sum of the [SII] doublet with the [SIII] lines 
at 9069 and 9532\AA\, normalized to the strength of H$\beta$.

	S$_{23}$ can not be computed from the GMOS data as H$\beta$ was not observed. 
In lieu of S$_{23}$, the sum of [SII]6700 and [SIII]9069 divided by H$\alpha$ 
is shown in Figures 4 -- 6. This ratio depends on metallicity and the ionization state, 
and so it is not surprising that it varies with location in the area observed with 
GMOS. The [SII]$+$[SIII]/H$\alpha$ ratios in HB, H4, H9, HA, and H2 are remarkably 
similar, arguing that these nebulae likely have similar [S/H] and ionization 
conditions. An exception is HC, where [SII]$+$[SIII]/H$\alpha$ is one half that 
in the other nebulae, and it should be recalled that the HeI7065/H$\alpha$ 
ratio in HC is also distinct. Of the star-forming regions 
considered here, HC has the largest projected distance 
from the center of NGC 55, and so might be expected to have a slightly lower 
mean metallicity than star-forming regions with smaller projected distances from 
the galaxy center. However, to the extent that the gas in the NGC 55 plane 
is well-mixed so that there are no major abundance variations, then the 
line ratios in HC more likely point to ionization conditions in that nebula that 
differ from those in the other star-forming regions.

\subsection{The Spectra of Individual Star-Forming Regions}

	Spectra of the H2 and H4 star-forming complexes, 
constructed by averaging signal in areas where H$\alpha$ emission is at least 
one-half of the peak value detected with GMOS, 
are shown in Figure 7. These spectra thus sample areas with 
the highest concentration of young stars, and so provide estimates for when the most 
recent large-scale episode of star formation occured in each complex. 
As NGC 55 is viewed almost edge-on, there is a contribution to the light from stars 
and gas that are not part of these star-forming regions. 
The contribution from such contaminating light was estimated from areas that are 
adjacent to each complex, and the mean spectrum constructed from these measurements was 
subtracted from the H2 and H4 spectra. As Figures 4 -- 6 demonstrate, 
emission line ratios are not spatially uniform throughout this part of NGC 55, and this 
introduces obvious uncertainties when correcting for background light. However, 
the background signal tends to be modest when compared with the light from the much 
brighter star-forming complexes.

\begin{figure*}
\figurenum{7}
\epsscale{1.0}
\plotone{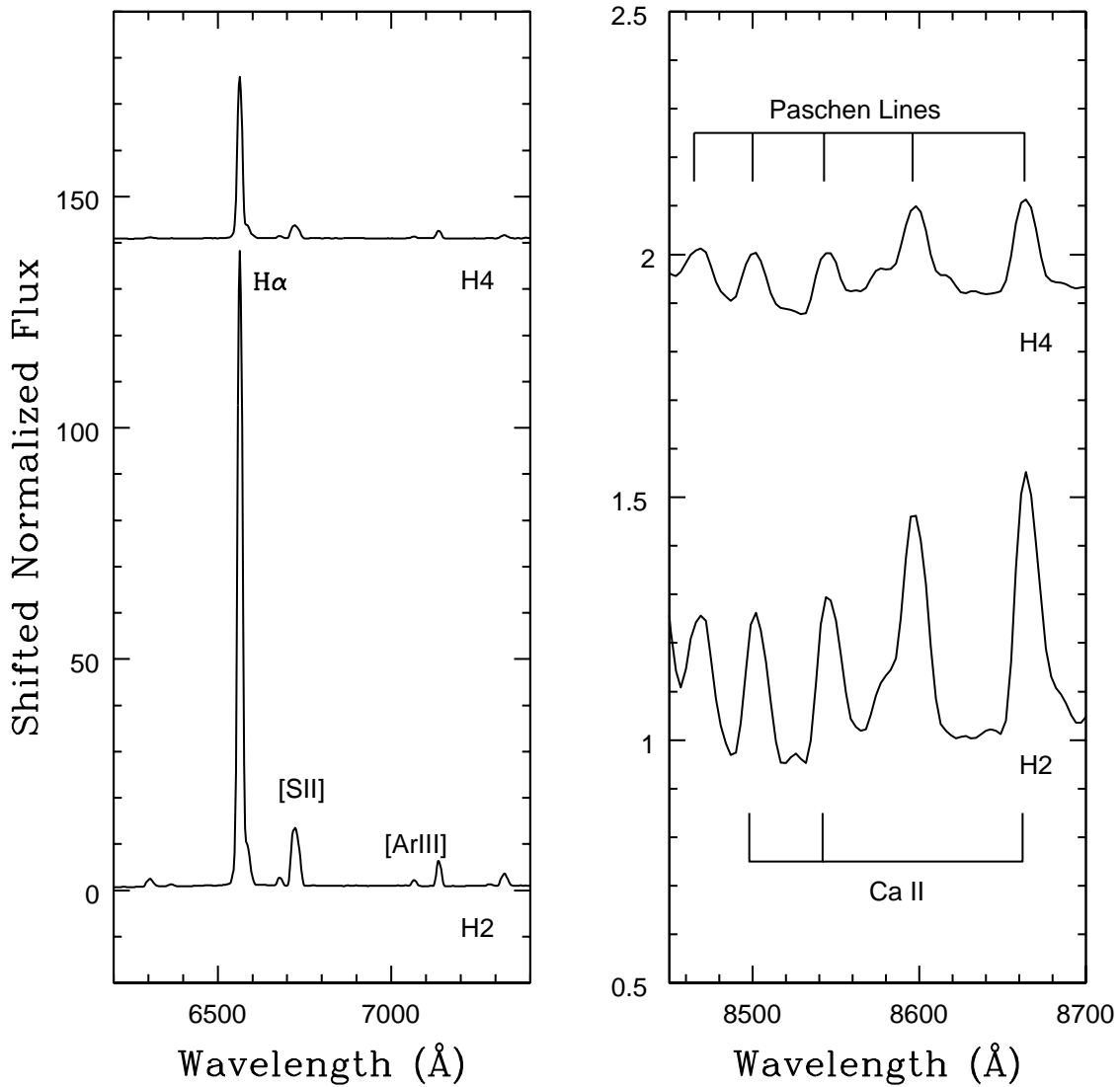}
\caption{Spectra of H2 and H4, constructed 
by combining signal in areas where H$\alpha$ emission is at least one 
half of the peak value in the area observed with GMOS. 
Wavelengths near H$\alpha$ and the near-infrared Ca triplet are shown. 
The spectra have been normalized to the continuum, and corrected 
for background light by subtracting the mean spectra of adjacent areas. 
The H4 spectra have been shifted vertically for display purposes. Emission lines in H2 
are more pronounced than in H4, suggesting that H2 contains hotter, 
more massive stars than H4. Wavelengths that sample Ca II lines are 
contaminated by Paschen series emission, complicating the 
detection of Ca absorption from luminous red stars.}
\end{figure*}

	Emission lines in H2 have larger equivalent widths than in H4. This 
is likely an age effect, with H2 containing hotter, more massive (and 
presumably younger) ionizing stars than in H4. 
The equivalent width of H$\alpha$ in H2 is $\sim 1700\AA$ while in H4 
it is $\sim 560\AA$. This places H2 on the flat part of the relation 
between the equivalent width of H$\alpha$ and time since the onset 
of star formation shown in Figure 83 of Leitherer et al. (1999), with an 
age log(t$_{yr}) \leq 6.4$, and a best estimate of log(t$_{yr}) \sim 6.2$. 
In contrast, H4 is on the rapidly descending part 
of this relation, with an age log(t$_{yr}) \sim 6.7$. Therefore, 
assuming that H2 and H4 are simple stellar systems 
that experienced an instantaneous burst of star formation and that each have 
Z=0.008, then the last major episode of star formation in H2 was within the past 2.5 Myr, 
whereas in H4 it was $\sim 5$ Myr in the past. 

	We caution that the age estimates made in the 
previous paragraph are based on spectra of the brightest part of the nebulae, and so 
probe mainly the most recent large-scale episode of star formation. 
Sub-structuring in star-forming complexes might 
be expected as the location of star formation within a giant molecular cloud 
may shift with time. In fact, compact HII regions are found near the edge of H2 
(Section 5), indicating that star formation is not centrally concentrated in that complex.
Large star-forming complexes may thus contain stars that formed 
over timescales of a few tens of Myr (e.g. De Marchi et al. 2011; 2017), which is 
comparable to the disruption time scale of giant molecular clouds (e.g. Murray 2011). 
The young ages found for H2 and H4 based on the H$\alpha$ equivalent widths do 
not preclude on-going star formation in these complexes, as the emission line 
spectra of systems experiencing continuous star formation reach an equilibrium 
value of a few Myr after the onset of star formation (Byler et al. 2017). 

	RSGs appear $\sim 6 - 8$ Myr after the onset of star 
formation, and spectroscopic signatures from such stars would then provide constraints on 
the duration of star formation. The Ca triplet lines are the deepest 
absorption features at red wavelengths in the spectra of RSGs, and are prominent features 
in the spectra of simple stellar systems with ages between $\sim 6 - 25$ Myr (e.g. 
Figure 97 of Leitherer et al. 1999). Unfortunately, emission lines from the Paschen 
sequence dominate the part of the spectrum that contains the Ca triplet, which is  
shown in the right hand panel of Figure 7.

	The challenges imposed by nebular emission aside, 
older stellar generations will likely develop a more diffuse 
spatial distribution than that of the youngest stars 
(e.g. De Marchi et al. 2017). Spectroscopic signatures of RSGs might then be seen in 
areas close to H2 and H4 where line emission is weaker. In Section 4.3 it is 
shown that there are areas adjacent to H4 where deep Ca triplet lines are seen.

	The spectra of other prominent H$\alpha$ sources 
in Figures 5 and 6, which we name HA, HB, and HC 
are compared in Figure 8. HA is at the center of the ring-shaped 
structure in the [SII]/H$\alpha$ and HeI/H$\alpha$ distributions 
that is to the immediate right of the minor axis of NGC 55. The 
equivalent width of H$\alpha$ in HA is comparable to that in H4, and the Paschen 
emission lines at wavelengths near 8600\AA\ have amplitudes like those in H4, 
suggesting similarities among the stars that power photoionization. 

\begin{figure*}
\figurenum{8}
\epsscale{1.0}
\plotone{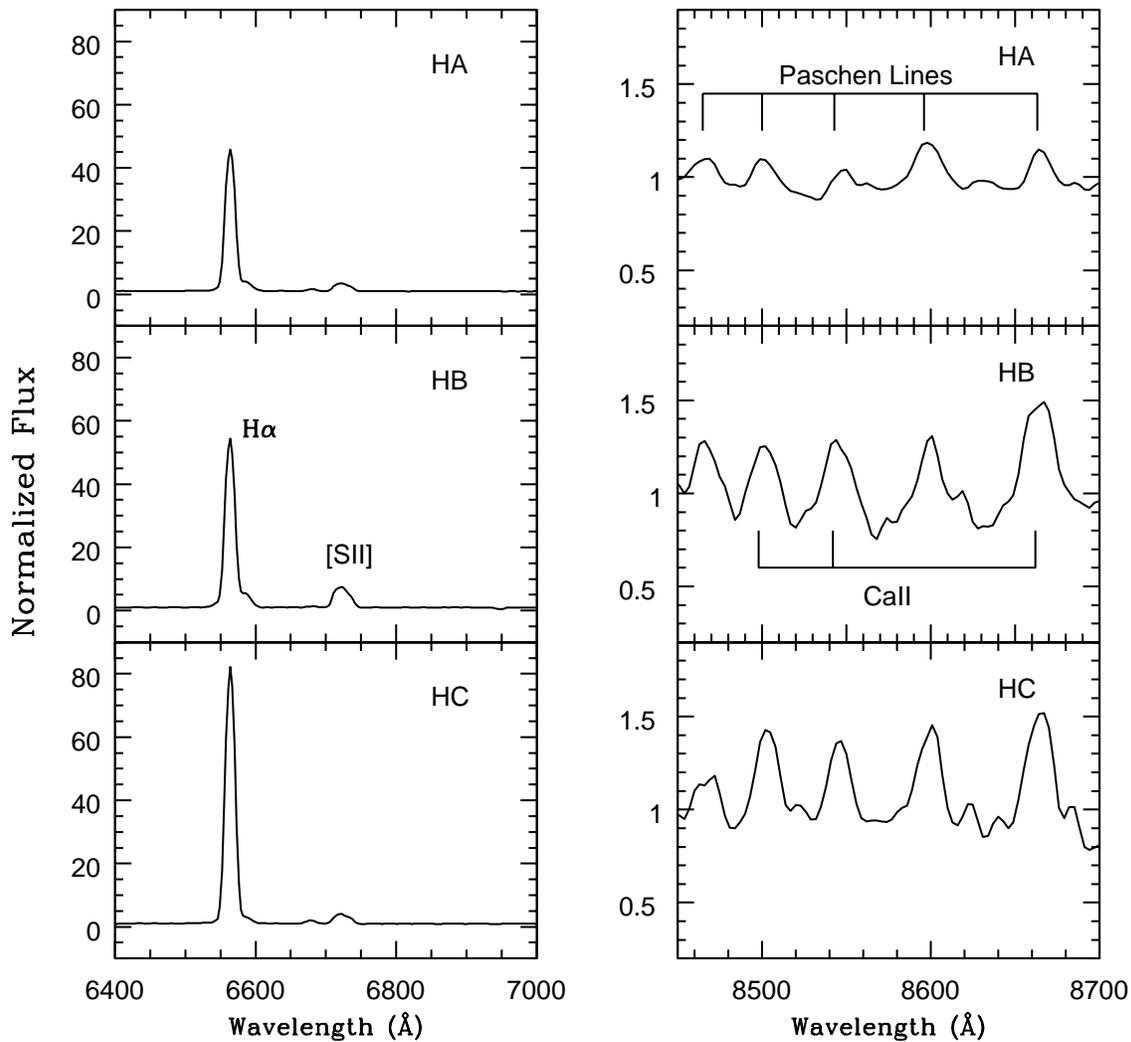}
\caption{Same as Figure 7, but showing spectra of the HA, HB, and HC star-forming areas. 
Emission lines from the Paschen series dominate the wavelength 
interval that contains the near-infrared Ca triplet.}
\end{figure*}

	HB and HC are prominent sources of H$\alpha$ emission in Figure 6, 
and the latter has a projected distance of $\sim 1$ kpc from the center of NGC 55. The 
FUV--NUV colors shown in Figure 2 indicate that HC is the furthest area 
of large-scale star formation from the center of NGC 55 in the north west portion 
of the disk. The equivalent width of H$\alpha$ in HB and HC is slightly 
larger than in HA and H4, but is weaker than in H2. The Paschen emission 
lines in HB are more pronounced than in HA and H4, and this could indicate that the 
ionization state of HB differs from that in HA or H4, possibly due to differences 
in the mix of stellar types that produce the ionizing radiation. 
[SII] is also stronger in HB than in either HA or HC, but 
[SII]/H$\alpha$ in HB is smaller than in H2. The equivalent width of H$\alpha$ in HC is 
smaller than in H2, although the Paschen emission lines in HC and H2 have similar 
equivalent widths.

\subsection{Absorption Features}

	The Ca triplet and the CN band head near 7900\AA\ 
are prominent absorption features in the spectra of evolved late-type stars. 
The spatial behaviour of the indices defined in Table 2 
that measure the depths of these absorption features 
is examined in Figure 9. Low signal levels prevent the measurement of the 
Ca3 and CN7900 indices near the ends of the GMOS slit, and so only the area 
near the center of NGC 55 -- covering the same spatial interval 
as in Figure 5 -- is shown. As the CN7900 and Ca3 indices 
probe features in the spectra of cool stars, the corresponding 
portion of the SPITZER [3.6] image is also shown in Figure 9. The H$\alpha$ 
measurements from Figure 5 are also repeated there.

\begin{figure*}
\figurenum{9}
\epsscale{1.0}
\plotone{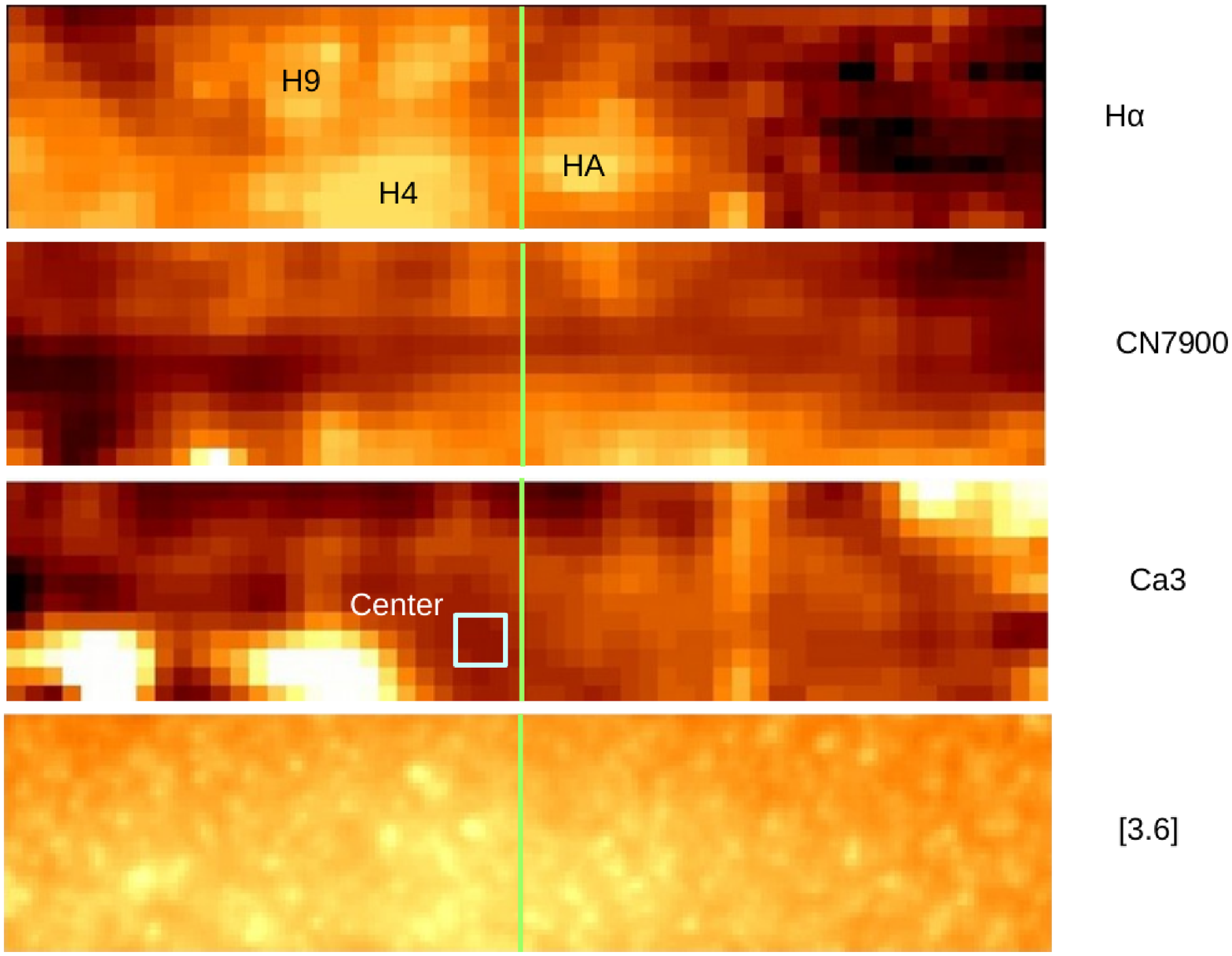}
\caption{Distribution of H$\alpha$ (top panel), CN7900 (second panel) and 
Ca3 (third panel) indices in the spatial interval examined in Figure 5. 
The indices are displayed with a logarithmic stretch with darker shades 
corresponding to deeper features. The bottom panel shows this same area imaged in 
[3.6]. The minor axis of NGC 55 is indicated with a green line. The distribution of CN7900 
and Ca3 indices shows structure that tends to be anticorrelated with H$\alpha$, in 
the sense that weak Ca3 absorption is found in areas with strong H$\alpha$ emission. 
Still, there are exceptions. An area with deep Ca3 
absorption is seen to the right of H4 near the minor axis of 
NGC 55, and this corresponds to a concentration of luminous sources in the [3.6] image. 
Other areas with deeper than average Ca absorption are also seen 
above and to the left of H4. It is evident from the [3.6] image that the stellar content 
in this part of NGC 55 is not uniformly mixed, and that there is a 
population of luminous evolved stars close to the center of NGC 55. The blue square 
indicates the area where the Center spectrum plotted in Figure 10 was extracted.}
\end{figure*}

	If the stars that dominate the near-infrared light 
are uniformly mixed throughout the central regions of NGC 55 then any changes in the 
projected distribution of the CN7900 and Ca3 indices would be due to non-uniform 
veiling by continuum emission. Paschen emission lines also overlap with 
lines of the Ca triplet. If continuum and line emission played 
a significant role in defining the depths of absorption features then the depths of 
absorption features and the equivalent width of H$\alpha$ emission would be coupled.

	The area shown in Figure 9 covers a range of H$\alpha$ equivalent widths, and 
there is a tendency for the areas of weakest Ca3 and CN7900 to occur 
where H$\alpha$ emission is strongest. Still, the areas of deepest Ca 
absorption are not exclusively found where the equivalent width of H$\alpha$ is 
smallest, but instead tend to occur between H4 and 
HA, near the minor axis of NGC 55. There are other areas with 
comparatively deep CN7900 and Ca triplet absorption in 
Figure 9, indicating that the stars that produce the deep absorption features are not 
restricted to the area near the minor axis, but are in what may 
be clusters or associations located throughout the area shown in Figure 9. 

	There is a concentration of luminous red stars in 
the [3.6] image in the lower panel of Figure 9 that coincides with 
the pocket of deep Ca triplet absorption between H4 and HA. The mean spectrum of the area 
marked in Figure 9 is shown in Figure 10. The wavelength interval near H$\alpha$ is shown 
in the left hand panel, while the wavelength interval that includes the Ca lines 
is shown in the right hand panel. Also shown is 
the median of the continuum-corrected spectra of the entire area sampled with GMOS. This 
spectrum notionally shows the `typical' spectroscopic characteristics in this 
part of NGC 55.

	The equivalent widths of H$\alpha$ emission in the 
two spectra in Figure 10 are comparable. The similarity in H$\alpha$ equivalent widths 
notwithstanding, the Ca triplet lines in the Center 
spectrum are deeper than in the median spectrum. The deep Ca absorption is 
attributed to the luminous red stars -- which likely are RSGs -- in 
the [3.6] image. The pockets of deep CN7900 absorption 
near the left hand edge of Figure 9 where CN7900 is very strong 
are not associated with obvious stellar groupings in the [3.6] image.

\begin{figure*}
\figurenum{10}
\epsscale{1.0}
\plotone{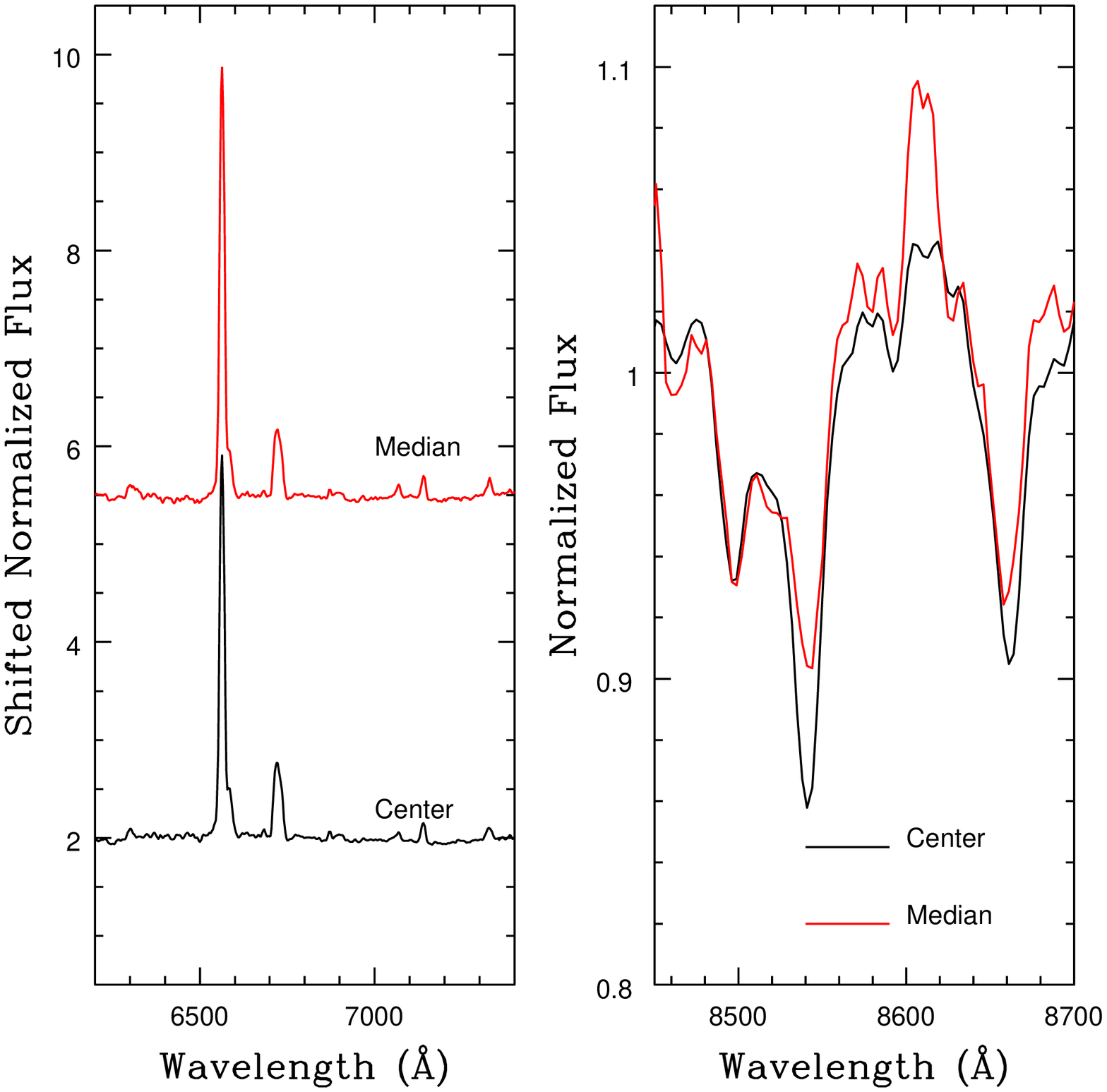}
\caption{Mean spectrum of the area marked `Center' in Figure 9, and 
the median spectrum of the entire area sampled with 
GMOS (`Median'). The spectra have been normalized to the continuum, and those in the 
left hand panel have been shifted vertically for display purposes. 
H$\alpha$ emission in the Center spectrum is only slightly weaker 
than that in the Median spectrum. However, the Ca lines in the Center spectrum 
are much deeper than in the Median spectrum.}
\end{figure*}

	There is an absence of bright red stars in the [3.6] image in the main bodies of 
H4, H9, and HA, and this can be used to set an upper limit to the time since the onset of 
star formation in these locations. It is unlikely that luminous red 
stars are present but obscured by dust in these areas, as gas tends to be 
ejected from star clusters on timescales that are comparable to the onset of 
supernovae, which occur a few Myr after the start of star formation 
(e.g. Hollyhead et al. 2015). For comparison, the RSG phase of evolution 
starts 6 -- 8 Myr after the initiation of star formation, and so this 
evolutionary phase is expected to occur after gas is ejected. The absence of 
RSGs suggests that the areas of present-day large-scale star formation in Figure 9 
have been forming stars for timescales of no more than a few Myr. 

	While H4, H9, and HA may be relatively young structures, the presence of RSGs 
throughout the central regions of NGC 55 indicates that large scale star formation has 
been on-going for at least many Myr. In fact, the spatial distribution of RSGs in the [3.6] 
image is clumpy, suggesting that there are numerous fossil star-forming regions near the 
center of NGC 55. A large population of luminous young red stars near 
the center of NGC 55 is not unexpected. Engelbracht et al. 
(2004) use MIPS images to examine the star-forming activity 
in NGC 55, and find that roughly one-third of the flux at 24$\mu$m originates from 
compact regions, almost all of which are near the center of the galaxy.
They estimate a global star formation rate (SFR) for NGC 55 of 0.22 M$_\odot$ year$^{-1}$, 
and the SFR in compact regions is then $\sim 0.07$M$_\odot$ year$^{-1}$. 
If star formation in this part of NGC 55 proceeded at a constant rate for 10$^7$ 
years -- which is the timescale consistent with the presence of RSGs -- 
then $\sim$10$^6$ solar masses of stars would have been produced assuming a 
solar neighborhood-like mass function. This corresponds to $\sim 50$ clusters similar in 
size to The Arches, which has a mass of $2 \times 10^4$ M$_{\odot}$ 
(Espinoza et al. 2009). A larger mass of stars would have formed if 
star formation proceeded continuously near the center of NGC 55 at this rate for more 
than 10$^7$ years.

\section{POINT SOURCES}

	A number of bright point sources are present in the GMOS observations, 
and the spectroscopic properties of these are discussed in this section. 
It is unlikely that the objects found here are individual stars given the high stellar 
density near the center of NGC 55, coupled with the $\sim 1.5$ arcsec 
FWHM image quality of these data. Instead, many -- if not all -- 
are undoubtedly either asterisms, compact nebulae, and/or compact star clusters that 
appear as a bright single object (e.g. Figure 15 of Stephens et al. 2003). 

	Sources were identified by eye in individual disk spectra. 
This was done because the 2 arcsec wide slit makes the detection of individual 
objects with sizes of 1 - 1.5 arcsec FWHM in mosaiced images problematic -- 
meaningful angular resolution measurements on sub-arcsec scales are only available 
in the direction along the spectrograph slit. Emphasis was also placed on relatively bright 
point sources so as to obtain spectra with moderately high S/N ratios, 
and so an automated detection algorithm was 
deemed to be unnecessary. Source identification was performed after collapsing the 
spectrum of each offset along the wavelength axis in the 7000 -- 
8500\AA\ interval to approximate the bandpass of the $i'$ filter. 
The use of light over a wide wavelength range 
suppresses (but does not completely remove) a bias in favor of emission line objects. 
The success of removing this bias is borne out in the types of objects that are 
detected, as only some have prominent emission lines in their spectra. 

	Spectra of point sources were extracted by summing 
the signal within the FWHM of the seeing disk. Sources that are obvious 
blends were not considered unless the peak signal from the contaminating source 
was less than half that of the primary target, 
and was separated from that source by at least one FWHM. 
Spectra to monitor local background light were extracted in the area surrounding target 
objects, and the mean of these was subtracted from each 
extracted spectrum. Given the crowded nature of the central regions of NGC 55 and 
the non-uniform spatial distribution of ionization characteristics then there are cases 
in which a reliable local sky spectrum could not be obtained, and there is evidence of 
over- or under-subtraction of emission lines in some spectra.

	Photometric measurements of these objects were 
obtained from the [3.6] and [4.5] images discussed in Section 3. 
The angular resolution of the SPITZER images is not greatly different from that 
of the GMOS data, simplifying the task of matching targets in the two datasets. 
Photometry in these filters provides information about the nature of 
the sources, especially those in which the light is dominated by highly 
evolved cool stars. 

	The photometric measurements were made with the point spread function 
(PSF)-fitting routine ALLSTAR (Stetson \& Harris 1988). 
A PSF for each filter was constructed from isolated bright stars, 
and routines in the DAOPHOT (Stetson 1987) package were used to obtain a source list 
and preliminary aperture photometry measurements. An object was considered to be 
detected only if it was identified in both the [3.6] and [4.5] images. 
The photometry was calibrated by applying the zeropoints measured by Reach et al. (2005). 
Only photometric measurements of point sources with spectra are discussed here; 
the photometry of a much larger sample of sources in and around NGC 55 detected 
in the SPITZER images will be presented in a separate paper (Davidge 2019, in preparation).

	Sources were sorted by eye into six groups based on their spectroscopic 
properties: (1) HII regions, (2) early-type stars, (3) M giants/supergiants, 
(4) C stars, (5) composite systems that show a mix of spectroscopic signatures, 
and (6) objects with largely featureless spectra. Objects in the last two groups have 
no clear spectroscopic type and are not discussed further as they 
may be the result of unreliable sky subtraction. The locations of the objects in the first 
four groups are shown in Figure 11. The panels in this figure cover the same areas 
as those in Figures 4 -- 6. 

\begin{figure*}
\figurenum{11}
\epsscale{0.9}
\plotone{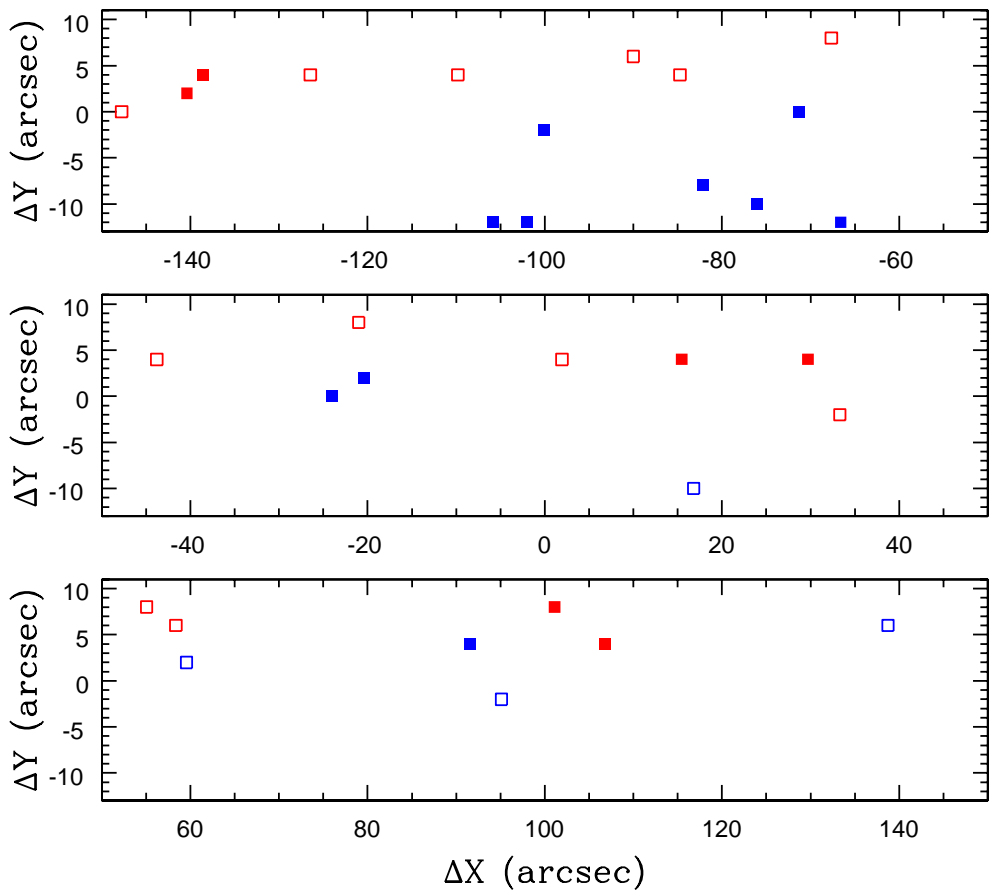}
\caption{Locations of individual sources, with co-ordinates measured along the major 
and minor axes. The intersection of the major and minor axes define the 
origin of the co-ordinate system. The panels cover the same areas as in Figures 4 -- 6, 
and the co-ordinates are listed in the second and 
third columns of Table 3. Sources with spectra classified as M-type 
(open red squares), C star-type (filled red squares), early-type (open blue squares), and 
HII region-like (filled blue squares) are shown. The majority of 
compact HII regions are located near H2 and H4, suggesting that they 
may be the result of star formation propogating throughout these areas.
The remaining sources are distributed more-or-less uniformly over the northern 
part of the area observed with GMOS. This is an area where line emission is less intense 
and obscuration due to dust is lower than in the southern half of the area studied. 
The image quality in some of the spectra that sample the southern half of the field 
is also slightly poorer than in the northern half, although the difference in 
image quality is small enough so as not to markedly affect the detection of point 
sources.}
\end{figure*}

	Magnitudes and locations of the 
sources are shown in Table 3. $\Delta X$ and $\Delta Y$ refer 
to offsets in arcsec from the center of NGC 55, with the X axis paralleling the 
major axis of NGC 55. The sources are not distributed uniformly 
across the field, with the majority of objects that have star-like spectra 
located north of the major axis. This positional discrepancy is likely due to 
a number of factors. The line emission and amount of dust 
obscuration is greater in the southern half of the 
area observed that has $\Delta X > 0$ than in the northern half of the field. 
In addition, the image quality of some of the spectra 
in the southern half of the field is slightly poorer than in the northern half, 
although the difference is no more than a few tenths of an arcsec and is not expected 
to hinder greatly the detection of bright sources. In fact, most of the compact HII regions 
are found in the southern half of the GMOS field.

\begin{deluxetable*}{llcccccc}
\tablecaption{Sources with Angular Sizes Compable to the Seeing Disk}
\startdata
\tableline\tableline
ID & Type & $\Delta X$ & $\Delta Y$ & RA & Dec & [4.5] & [3.6]--[4.5] \\
 & & (arcsec) & (arcsec) & (2000) & (2000) & & \\
\tableline
p4-4 & HII & (74.4) & (3.6) & (00:14:46.8) & (--39:11:23.2) & -- & -- \\
p2-7 & HII & -37.8 & 4.2 & 00:14:58.1 & -39:12:10.3 & 14.16 & 0.39 \\
c-5 & HII & -40.2 & -1.8 & 00:14:58.2 & --39:12:18.7 & 16.14 & -0.40 \\
c-6 & HII & (-88.2) & (-0.6) & (00:15:03.1) & (--39:12:36.8) & -- & -- \\
m2-5 & HII & -114.6 & 0.0 & 00:15:05.7 & --39:12:48.2 & 15.13 & 0.70 \\
m8-3 & HII & (-99.6) & (-9.0) & (00:15:03.9) & (--39:12:51.6) & -- & -- \\
m10-5 & HII & -91.8 & -9.6 & 00:15:03.1 & --39:12:49.7 & 16.18 & -0.12 \\
m12-1 & HII & -82.8 & -13.2 & 00:15:02.0 & --39:12:50.3 & 16.30 & -0.50 \\
m12-2 & HII & (-119.4) & (-12.6) & (00:15:05.8) & (--39:13:04.4) & -- & -- \\
m12-3 & HII & (-123) & (-12.0) & (00:15:06.1) & (--39:13:05.6) & -- & -- \\
 & & & & & & & \\
p8-2 & M & 34.8 & 5.4 & 00:14:50.8 & --39:11:37.8 & 14.87 & 0.26 \\
p8-5 & M & (-84.6) & (7.8) & (00:15:03.1) & (--39:12:27.6) & -- & -- \\
p6-6 & M & (-107.4) & (6.0) & (00:15:05.1) & (--39:12:39.7) & -- & -- \\
p4-9 & M & -13.8 & 2.4 & 00:14:55.6 & --39:12:02.0 & 15.47 & 0.60 \\
c-10 & M & -165.6 & -2.4 & 00:15:10.8 & --39:13:11.9 & 16.61 & -0.25 \\
m2-3 & M & 17.4 & -1.2 & 00:14:52.4 & --39:11:53.1 & 15.69 & 0.17 \\
p8-3 & M & -37.8 & 7.2 & 00:14:58.0 & --39:12:04.1 & 16.63 & -0.09 \\
p6-3 & M & 39.0 & 6.6 & 00:14:50.5 & --39:11:33.8 & 16.98 & -0.11 \\
p4-10 & M & -60.0 & 3.6 & 00:15:00.4 & --39:12:20.5 & 15.11 & 0.27 \\
p4-13 & M & -100.2 & 4.8 & 00:15:04.4 & --39:12:36.3 & 16.30 & 0.01 \\
p4-14 & M & -123.6 & 4.2 & 00:15:06.8 & --39:12:47.1 & 16.49 & 0.18 \\
p4-15 & M & -140.4 & 3.6 & 00:15:08.4 & --39:12:57.8 & 16.04 & -0.03 \\
 & & & & & & & \\
p8-1 & C & 85.2 & 10.2 & 00:14:45.9 & --39:11:10.8 & 15.28 & 0.60 \\
p4-1 & C & 90.0 & 3.0 & 00:14:45.2 & --39:11:17.9 & 14.92 & 0.58 \\
p4-5 & C & 11.4 & 2.4 & 00:14:53.1 & --39:11:50.8 & 14.73 & 0.45 \\
p4-7 & C & (-1.8) & (3.6) & (00:14:54.5) & (--39:11:55.7) & -- & -- \\
p4-16 & C & -154.8 & 6.6 & 00:15:10.2 & --39:12:54.0 & 16.57 & 0.36 \\
p2-12 & C & (-157.2) & (1.8) & (00:15:10.1) & (--39:13:04.3) & -- & -- \\
 & & & & & & & \\
p6-1 & OB & 178.2 & 5.4 & 00:14:36.4 & --39:10:36.5 & 16.57 & -0.06 \\
p2-4 & OB & (84.6) & (1.2) & (00:14:45.7) & (--39:11:20.8) & -- & -- \\
m2-2 & OB & 126.0 & 177 & 00:14:41.4 & --39:11:08.5 & 15.91 & -0.47 \\
m10-4 & OB & 32.4 & 166 & 00:14:51.0 & --39:11:41.8 & 15.91 & 0.11 \\
\tableline
\enddata
\end{deluxetable*}

	There is the potential for ambiguity when matching objects detected at 
widely different wavelengths given that the light might originate from very different 
mechanisms at each wavelength (e.g. photospheric light in the GMOS observations, 
versus the potential for significant levels of thermal 
emission in the SPITZER data). Differences in the levels of 
extinction between the wavelengths sampled by GMOS and SPITZER further complicate 
source matching, as do differences in angular sampling. 
The observations for this program were defined using images from the 
Digital Sky Survey (DSS), and the DSS images were 
used to align the GMOS and SPITZER observations. Reference 
sources that were common to both datasets were identified. The alignment 
of the two datasets involved correcting for angle of rotation and differences in pixel 
sampling, followed by the application of offsets along the X and Y axes.

	A GMOS source was considered to be matched to a SPITZER counterpart if the 
co-ordinates agreed to within 3 arcsec. While 3 arcsec is a generous 
matching radius, it is only $1.5\times$ the FWHM of the SPITZER 
[3.6] observations and the width of the GMOS slit. Past experience indicates that a 
matching criterion of $1 - 2\times$ FWHM is sufficient to identify real source 
matches, while limiting the numbers of false matches. In any event, the majority of 
matched sources have co-ordinates that agree to much better than 3 arcsec. The 
co-ordinates of sources that did not have photometric detections, and so are based solely 
on GMOS-derived co-ordinates, are shown in brackets.

	The ([4.5],[3.6]-[4.5]) CMD of the spectroscopically-identified point sources 
is shown in Figure 12. Many of the objects scatter about [3.6]--[4.5] $\sim 0$, which 
is the locus of sources that have effective temperatures higher than a few thousand K. 
There is also a population of bright objects that have comparatively red 
[3.6]--[4.5] colors, and these are sources where thermal emission 
from hot dust likely contributes to the SED. The majority of sources are at least 
1 magnitude above the faint limit of the SPITZER observations, which is
indicated by the dashed line in Figure 12. This is not unexpected as the comparatively 
shallow GMOS observations will detect only relatively bright sources. 

\begin{figure*}
\figurenum{12}
\epsscale{1.0}
\plotone{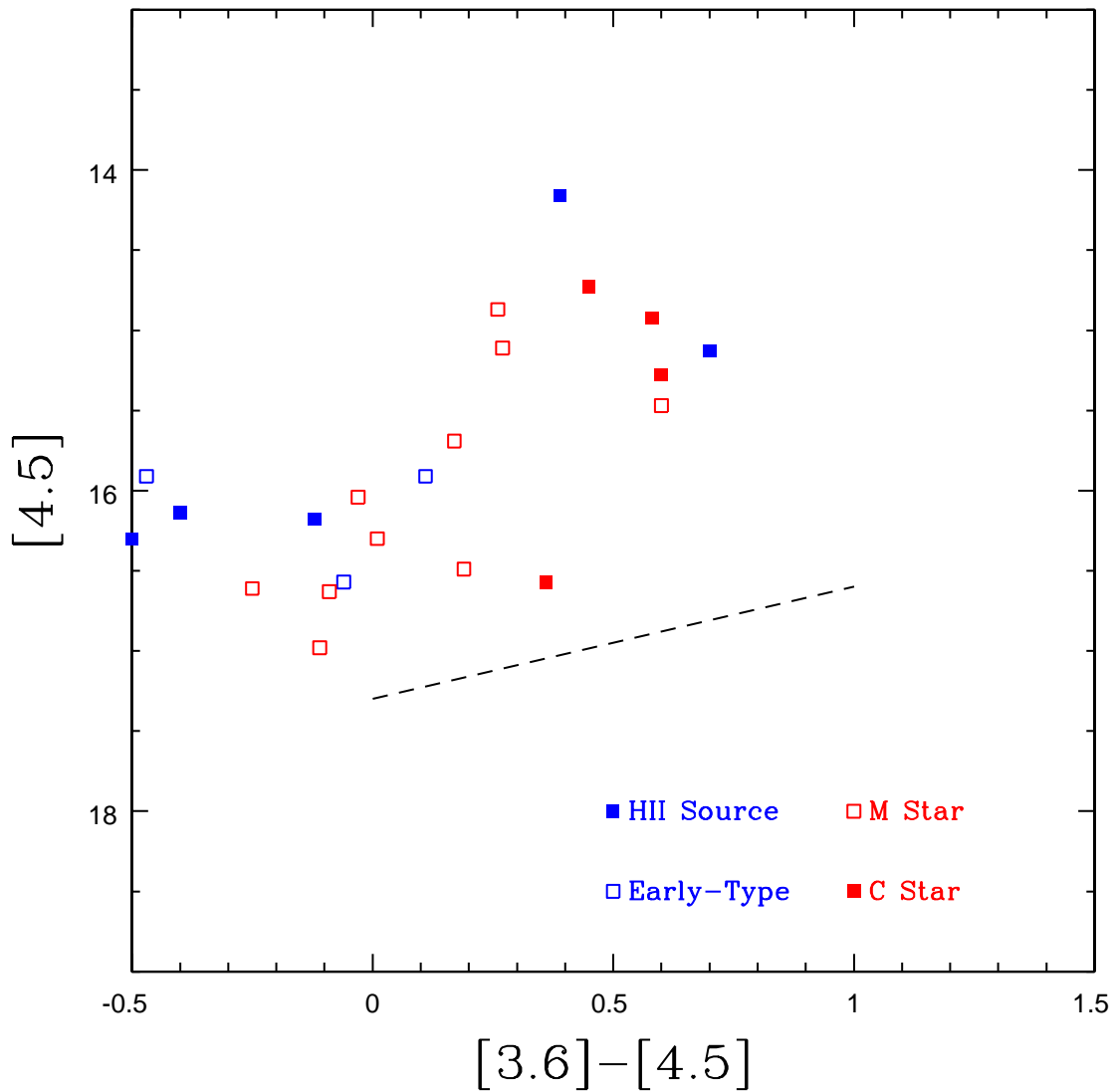}
\caption{The ([4.5],[3.6]-[4.5]) CMD of point sources in the GMOS spectra. 
The light from each of these objects likely comes from many 
stars, although a small number of objects may dominate the 
light output at these wavelengths. The dashed line 
shows the faint limit of the SPITZER observations. Objects in which the light 
originates from sources with effective temperatures in excess of a few thousand K 
have [3.6]--[4.5] $\sim 0$. The broad color distribution of sources 
that are matched with HII regions reflects the diverse SEDs of sources 
within star-forming areas. The two bright, red sources associated 
with HII regions are probably highly obscured young clusters.}
\end{figure*}

\subsection{Compact HII Regions}

	The spectra of sources with emission lines that might be indicative of HII regions 
are shown in Figure 13. As the sources discussed in this section were selected because they 
are compact, then the spectra of larger, more extended areas of H$\alpha$ emission 
in Figures 4 -- 6 are not included in this figure. With one exception, the compact HII 
regions are found near the outer boundaries of H2 and H4, and this spatial distribution 
suggests that these are pockets of localized star formation that are triggered by 
activity within the larger star-forming complexes. 

\begin{figure*}
\figurenum{13}
\epsscale{1.0}
\plotone{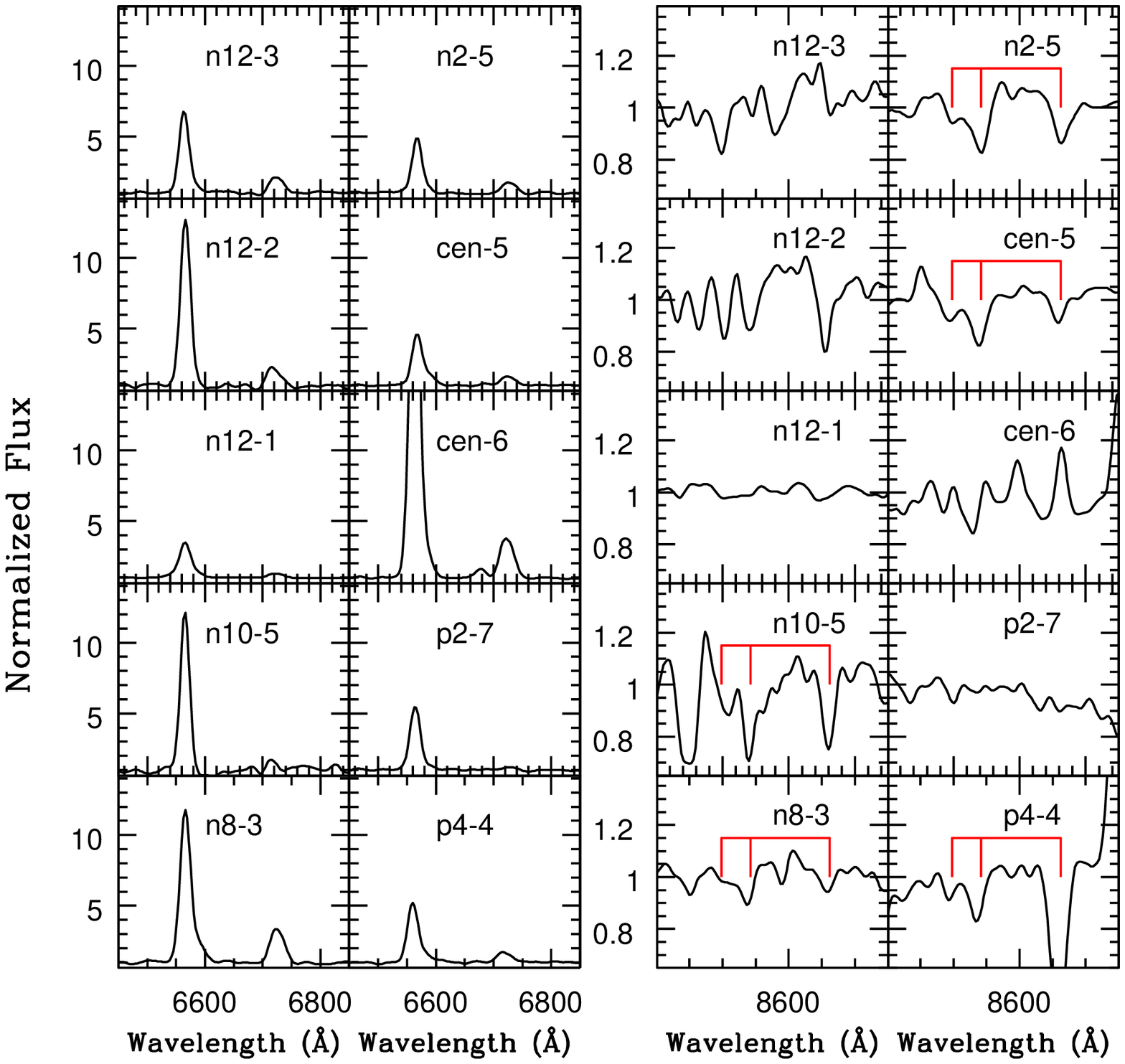}
\caption{Spectra of compact sources classified as HII regions. Wavelengths 
near H$\alpha$ are examined in the left hand panel, while wavelengths centered on the 
near-infrared Ca triplet are covered in the right hand panel. The 
spectra have been normalized to the continuum. All objects have 
[SII]/H$\alpha$ ratios that are consistent with photoionization generated close to 
the ionizing source(s); none are SNRs. H$\alpha$ emission is strongest in cen-6, and 
in many cases the emission lines in the Pa series are clearly seen in the 
wavelength interval containing the Ca triplet. Still, some of the spectra have 
prominent Ca triplet absorption, indicating that luminous red stars are concentrated 
within $\sim 10$ parsecs of the line-emitting regions. The wavelengths of the 
Ca triplet lines are marked in those spectra. The abnormally strong Ca line near 
8662\AA\ in the p4--4 spectrum is likely due to poor sky subtraction caused by 
non-uniformities in the background light.}
\end{figure*}

	The emission and absorption features in the Figure 13 spectra 
may be affected by the non-uniform distribution of line emission 
in the areas used to estimate background light levels. Keeping this caveat in mind, 
the typical ratio of [SII]/H$\alpha$ in Figure 13 is comparable to that in H2 and 
H4, as expected if the emission is due to photoionization that originates close to 
the ionizing stars. The [SII]/H$\alpha$ ratio also 
indicates that none of the objects are compact SNRs. 

	Paschen line emission is seen in the spectrum of cen--6, which also has the 
strongest H$\alpha$ emission line. However, Ca triplet absorption lines are also 
seen in some of the HII spectra, suggesting that luminous red 
stars -- presumably RSGs -- contribute significantly to the 
light. This absorption is most obvious in the spectra of n2-5 and cen--5. There is 
comparatively weak H$\alpha$ emission in the spectra of those sources, and so veiling 
from continuum emission and contamination from Pa emission 
lines is likely to be less of a factor than in sources with 
stronger H$\alpha$ emission. Still, the spectrum of p2--7 indicates that weak 
H$\alpha$ emission is not a guarantee that the Ca triplet will be detected. 
There thus appears to be source-to-source variations in the red stellar content that 
is the source of the Ca triplet lines. 

	What is the nature of the sources that produce Ca absorption 
in Figure 13? It is unlikely that the Ca lines originate 
from a widely dispersed, uniformly distributed population, as the 
signatures of such a population would be suppressed during background subtraction.
Still, the spatial resolution of the GMOS data at the distance of NGC 55 is 
$\sim 20 \times 20$ parcsecs, and so the stars that are the source of the Ca lines 
may not physically coincide with the HII region. Luminous red stars 
might be expected near large star-forming complexes 
if there has been star-forming activity for an extended period of time. 
If star formation has been on-going in n2--5 and 
cen--5 for at least 6 -- 8 Myr then the Ca absorption may originate from an 
older population that has had time to diffuse away from the present-day star-forming 
region. If these stars are free from areas of high extinction then 
the contrast between the Ca triplet lines and Pa line emission will be much more 
favorable for the detection of stellar absorption lines. 
Alternatively, RSGs and luminous AGB stars could originate 
in older clusters that have ages in excess of $\sim 8$ Myr and 
are located close to the HII regions. 

	Half of the HII regions were matched with sources in the SPITZER 
observations, and those that were matched 
have a broad range of [3.6]--[4.5] colors. The light from the objects 
detected by SPITZER likely originates from stars/star clusters within the HII region. 
These are expected to have a diverse range of SEDs that reflect differences 
in the amount of obscuration and the effective temperatures of the embedded sources.
The brightest spectroscopically-identified point source in [4.5] is associated with an 
HII region, and this is probably a highly 
obscured young cluster located inside the HII region. 
There is no correlation between [3.6]--[4.5] color 
and the detection of Ca triplet lines in the spectrum.

\subsection{Early and Late-Type Spectra}

	Spectra of objects with characteristics that are indicative 
of early-type stars are shown in Figure 14. The spectra of these 
sources have prominent Paschen series absorption lines, and 
the Paschen break near $0.82\mu$m is seen in all four spectra.
Whereas the Paschen line characteristics of the four spectra 
in Figure 14 show good source-to-source agreement, this is not the case for 
H$\alpha$. While H$\alpha$ is in absorption in 
three spectra, it appears that the H$\alpha$ and [SII] emission lines in 
the spectrum of m10--4 have been over-subtracted. This illustrates the difficulty 
obtaining reliable H$\alpha$ strengths in areas with non-uniform line emission.

\begin{figure*}
\figurenum{14}
\epsscale{1.0}
\plotone{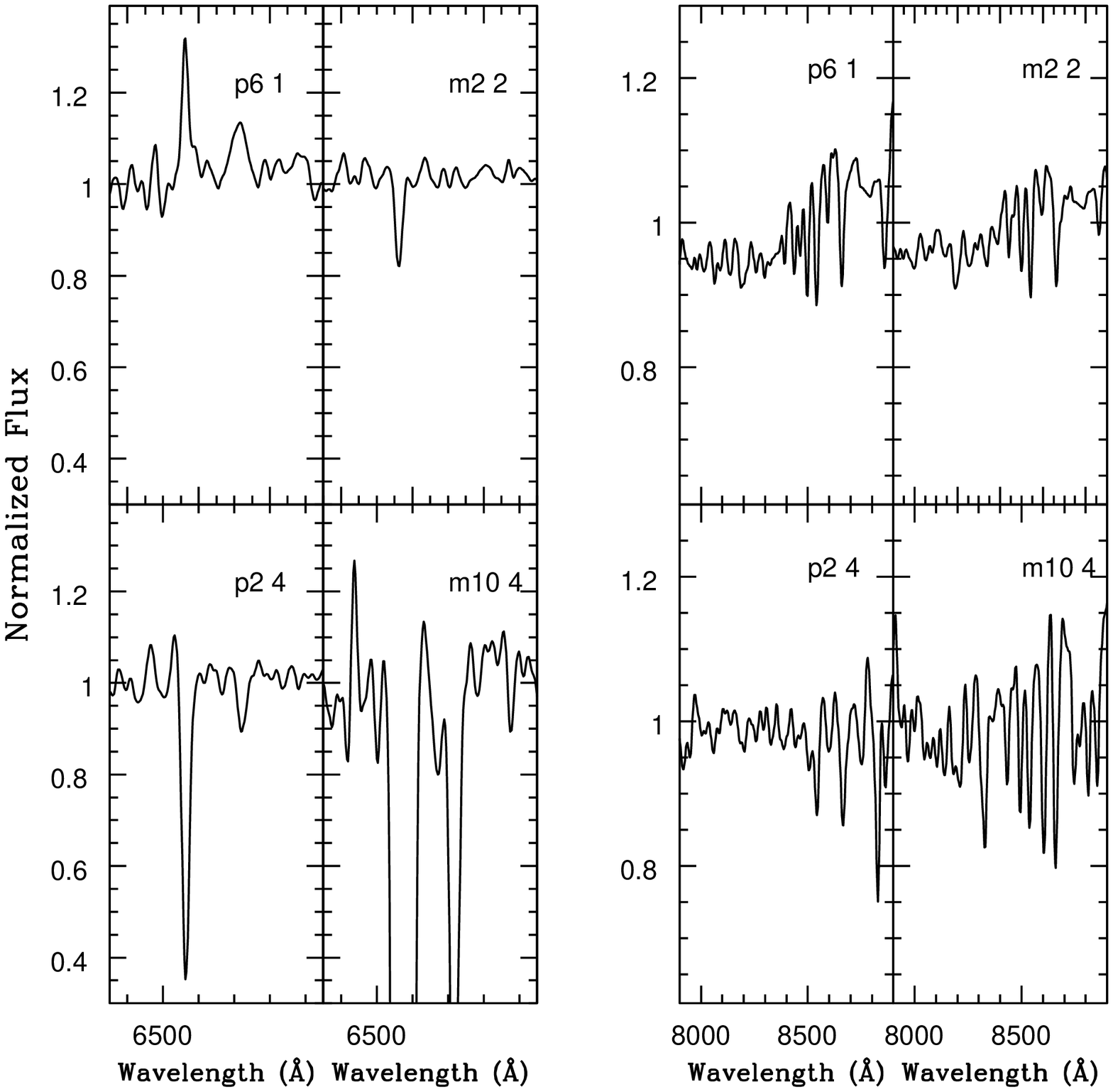}
\caption{Spectra of sources where the light is dominated 
by early-type stars. The spectra have been normalized to the continuum. Prominent Paschen 
line absorption is detected in all four sources, and the Paschen break is seen near 
8200\AA\ . With the exception of p6--1, these objects have H$\alpha$ in absorption. 
However, the spectra in the left hand panel illustrate how uncertainties in background 
subtraction can affect H$\alpha$ characteristics, as the extremely 
deep H$\alpha$ and [SII] lines in the spectrum of m10--4 are likely due to the 
over-subtraction of nebular emission lines. The depths of the H$\alpha$ absorption lines 
may also be affected by H$\alpha$ emission that is intrinsic to the sources.}
\end{figure*}

	The light from the objects with spectra in Figure 14 is likely dominated by 
a single bright blue supergiant (BSG) or a small number of BSGs. 
The dominant source of light is/are then objects that have ages of 
no more than a few tens of Myr. Three of the four objects 
with early-type spectroscopic characteristics were 
matched with SPITZER sources, and the [3.6]--[4.5] colors of these 
suggest that thermal emission does not contribute significantly to the infrared light. 
If red stars are present within 10 -- 20 parsecs of the BSGs -- as might be expected 
if the BSGs are in clusters or associations -- then those red stars do not 
have large-scale circumstellar envelopes, such as those that are found around the most 
luminous AGB stars. 

	Sources with early-type spectroscopic characteristics 
are restricted to $X > 15$ arcsec in Figure 11, suggesting that they 
do not have an obvious physical connection with the present-day activity in 
H2 and H4. Still, we caution that the distribution of point sources is likely affected by 
selection effects, as the detection of stars/compact clusters 
in the part of the GMOS observations with negative 
offsets in X and Y is complicated by line emission. BSGs in or near 
star-forming regions might also evade detection 
if they are obscured by dust along the line of sight.

	Spectra of sources with TiO bands longward of $0.71\mu$m are shown in Figure 15. 
The Ca triplet is also seen in the spectra of all twelve objects, and it 
is likely that the objects with spectra in this figure are star clusters 
where the red/near-infrared light is dominated by M supergiants. 
If this is correct then these sources have ages $\sim 10 - 100$ Myr.

	The majority of the sources with spectra in Figure 15 were detected in both [3.6] 
and [4.5], and these have a broad range of [3.6]--[4.5] colors. This is consistent with 
the MIR SEDs of Galactic M giants, which also span a range of [3.6]--[4.5] colors (e.g. 
Reiter et al. 2015). While sources with late-type spectroscopic features are found 
in the northern half of the area observed with GMOS, as with BSGs this distribution 
might be due to selection effects.

\begin{figure*}
\figurenum{15}
\epsscale{1.0}
\plotone{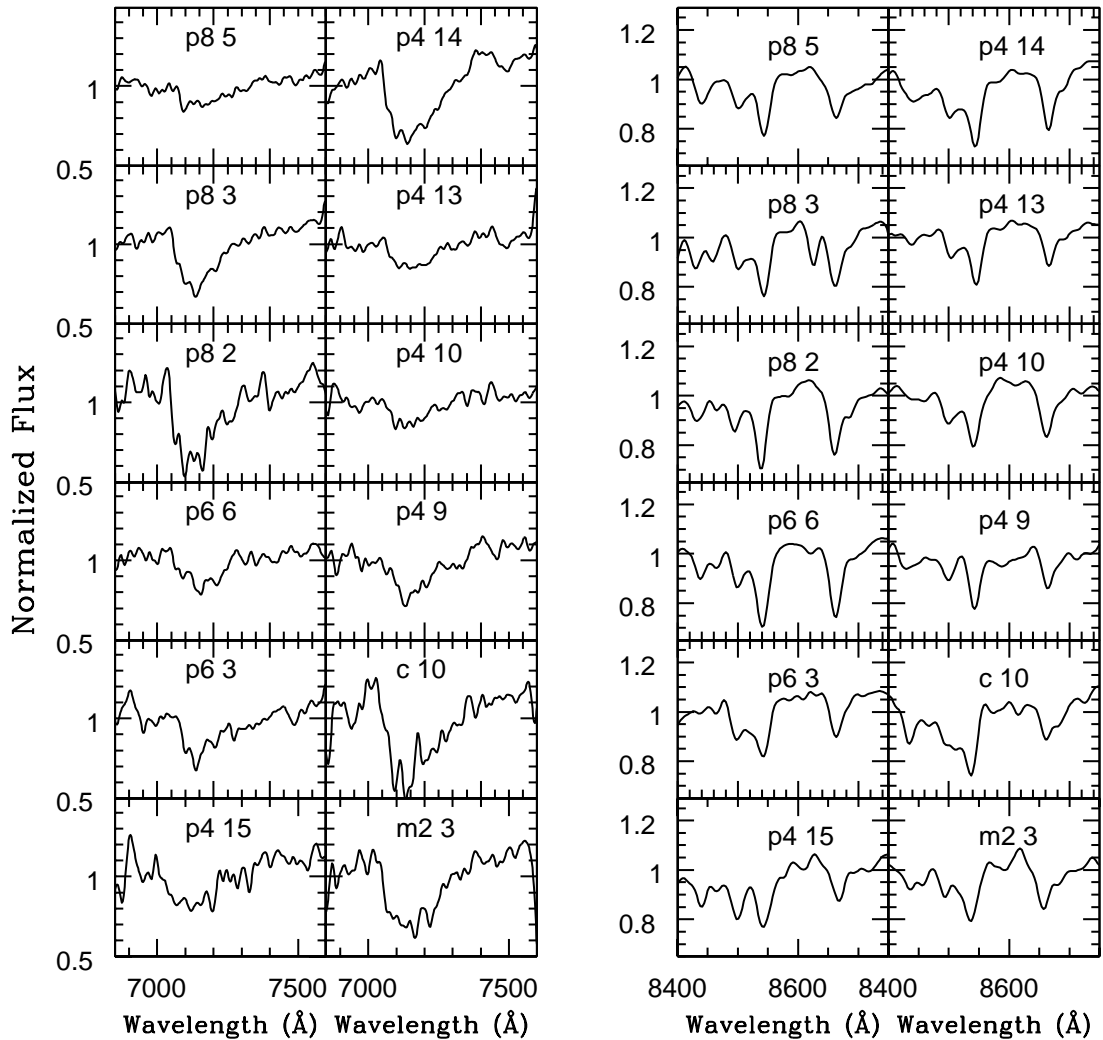}
\caption{Spectra of point sources that have deep TiO bandheads at 7100\AA\ , 
indicating that the light is dominated by M giants. The spectra have been normalized 
to the continuum. Absorption lines of the Ca triplet are also present. The 
light from these sources is likely dominated by one or more RSGs.}
\end{figure*}

\subsection{Sources with Broad CN bands}

	C stars have been detected in significant numbers in the outer regions of NGC 55 
(Pritchet et al. 1987), and are likely also present in the central regions of the galaxy. 
Individual C stars can be detected in the outer regions of NGC 55 
as the stellar density is low, so that there is minimal 
dilution of the CN bands by contaminating light. While crowding makes the detection 
of single stars challenging in dense environments like the central regions of NGC 55, 
luminous C stars produce distinct signatures in the integrated spectra of at 
least some intermediate age LMC clusters (Davidge 2018a), opening the possibility 
that these features might be detected even if the light from a C star is blended with 
that of other objects. Luminous AGB stars are also photometrically variable, and they may 
dominate the signal from a resolution element when near the peak of their 
light curves (Davidge et al. 2010). 

	Figure 16 shows the GMOS spectra of six unresolved sources 
that have broad CN bands. Four of these were detected 
in the Spitzer images, and these have [3.6]--[4.5] colors that are consistent with 
them being highly evolved, intrinsically luminous objects that have the circumstellar dust 
envelope that is found around many C stars. Also shown in 
Figure 16 are two spectra from Davidge (2018a): one is the 
spectrum of 2MASS05284449--6614309, which is the brightest C star in the LMC cluster 
NGC 1978, while the other is the spectrum of NGC 1978 Pointing 3 (P3). The latter 
is included to demonstrate how the integrated light from a cluster is affected by the 
presence of a C star. NGC 1978 has an age of 1.9 Gyr (Mucciarelli et al. 
2007), which is similar to the luminosity-weighted mean age estimated for 
NGC 55 at red/NIR wavelengths by Davidge (2018a). 

\begin{figure*}
\figurenum{16}
\epsscale{1.0}
\plotone{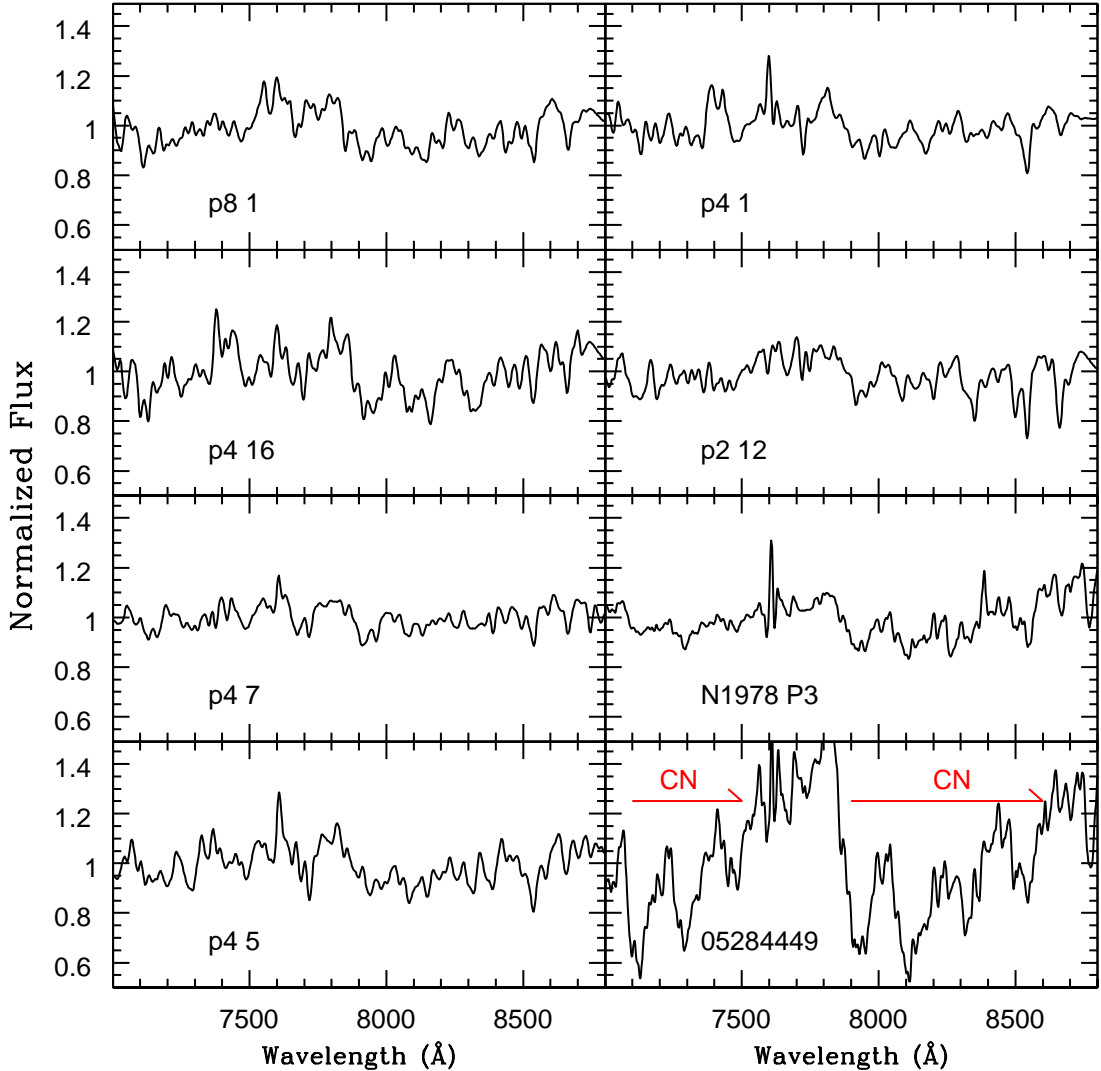}
\caption{Spectra normalized to the continuum of sources with CN absorption bands. Two 
spectra from Davidge (2018a) are also shown: (1) the spectrum of the C star 
2MASS05284449--6614309, which is the most luminous C star in NGC 1978, 
and (2) the spectrum of LMC cluster NGC 1978 Pointing 3 (P3), which shows the light of 
2MASS05284449--6614309 combined with that of the underlying cluster. A 
comparison of the spectra of NGC 1978 P3 and 2MASS05284449--6614309 indicates 
that while the CN bands that are signatures of C stars are subdued in integrated 
light, they can still be detected in the composite spectrum.
The depths of the CN bands in the NGC 55 sources 
suggest that the light originates from a stellar system, 
rather than a single star. The similarity between the spectra of the sources 
in NGC 55 and NGC 1978 P3 is likely a selection effect (see text).}
\end{figure*}

	The spectra of NGC 55 sources in Figure 16 have CN bands between 0.7 
and $0.9\mu$m that are similar in shape and depth to those in NGC 1978 
P3, and we suspect that these objects are compact 
intermediate age clusters with ages similar to that of NGC 1978. 
The spatial distribution of these objects is also consistent with them being AGB stars. 
Davidge (1998) found that luminous AGB stars in the southern half of 
NGC 55 are well-mixed over kpc spatial scales. 
The six NGC 55 sources in Figure 16 are distributed across the field sampled by 
GMOS, as expected for objects that have ages that allow them to disperse 
over large distances from their birth environments. 

	There is a remarkable similarity between the spectra of the NGC 55 sources 
in Figure 16 and NGC 1978 P3, and we suspect that 
this is largely a selection effect. The noise level in the GMOS spectra is 
such that clusters in which the C star contributes a much smaller fraction of 
the total signal than that from NGC 1978 will likely not have detectable CN bands. 
As for systems that might have stronger CN bands, it is anticipated that these will be 
rare in a crowded environment if observed with 
a $\sim 20$ parsec intrinsic resolution, as it 
would require a large number of C stars in one cluster or the fluke detection of an 
extremely luminous C star in a low mass cluster. The probability of either occurence 
is low given the rapid pace of C star evolution. 

\section{DISCUSSION \& SUMMARY}

	Long-slit spectra recorded with GMOS on GS that sample red and near-infrared 
wavelengths have been used to examine the projected distribution of emission and 
absorption features throughout the central regions of the nearby dwarf galaxy NGC 55. 
The results can be summarized as follows:

\noindent{1)} The central regions of many late-type galaxies show evidence 
for on-going or recent star formation, and NGC 55 is no exception. 
Much of the light from the H2 and H4 star-forming complexes 
at red wavelengths originates from line and continuum emission, with 
no evidence of Ca triplet absorption. The absence of Ca triplet absorption is consistent 
with the non-detection of luminous red stars in [3.6] images, suggesting that 
star formation in H2 and H4 has continued for only a few Myr. The spectra of three 
other large star-forming regions (HA, HB, and HC) have also been investigated. 
The equivalent widths of H$\alpha$ in these areas are comparable to 
those in H2 and H4, indicating that large star-forming areas in the disk of NGC 55 
extend out to projected distances of $\sim 1$ kpc from the galaxy center. 

\noindent{2)} Evidence is found for fossil star-forming regions near the center of NGC 55. 
Davidge (2005) discussed $J$ and $K$ observations of the 
center regions of NGC 55. Areas near H4, identified as `Cluster 1' by Davidge (2005), 
and H2, identified as `Cluster 2', contain a population 
of bright resolved RSGs, with ages $\sim 10$ Myr. A concentration of bright 
stars or star clusters are also seen near H4 and HA in the Spitzer [3.6] 
images in the lower panel of Figure 9. Similar concentrations of red 
stars are not obvious near HB and HC. The presence of bright red stars 
suggests that recent star formation within a few hundred 
kpc from the center of NGC 55 has been on-going for at least many Myr.

	There is a concentration of bright red sources in the [3.6] image 
near the intersection of the major and minor axes of NGC 55. These 
stars are distributed over $\sim 200$ parsecs, which is the approximate upper limit for the 
size of a GMC (Murray 2011). This central fossil star-forming region is part of 
an area with elevated light levels in the [3.6] luminosity 
profile in the lower panel of Figure 1. There is also localized 
deep Ca absorption in the integrated spectrum that is attributed to these stars. 
Even if many of these sources are blends or clusters, a concentration of very red 
compact clusters would also be a signature of recent large-scale star formation. 

	Westmeier et al. (2013) find non-circular motions near the center of NGC 55 
that they interpret as the signature of material that is channeled into the central 
regions by a bar. The centrally-concentrated population of luminous red stars and the 
evidence of on-going star formation in this area may then be the results of such 
a gas inflow. If the inflow of gas continues then it may affect the appearance 
of the center of the galaxy by assembling a moderately 
massive stellar nucleus or even bulge, assuming that the mass function is 
like that in the solar neighborhood. 

	That fossil star-forming regions are found near the center of NGC 55 is 
perhaps not surprising given that the current SFR in the central regions of NGC 55 is 
0.07M$_{\odot}$year$^{-1}$ (Section 4). This is similar to 
the SFR in the Central Molecular Zone (CMZ) of the Galaxy 
during at least the past few Myr (e.g. Barnes et a. 2017). The recent 
SFR in the CMZ has been sufficient to produce large clusters 
like The Arches and The Quintuplet. Similar rich stellar concentrations 
might then be expected near the center of NGC 55, and candidate structures are seen 
in the [3.6] image in Figure 9.
  
\noindent{3)} In addition to examining integrated light throughout the center of 
NGC 55, the spectra of objects that are point sources at angular 
resolutions of $\sim 1.5$ arcsec FWHM have also been examined. 
While a single bright object (or a handful of objects) might dominate the 
light from such sources, these "point" sources are almost certainly composite 
stellar systems. Some are compact HII regions, and these are potentially 
important as they may provide constraints on abundance gradients when compared with 
abundances measured from HII regions at larger radii. There is debate as 
to whether or not there is a metallicity gradient in the NGC 55 disk, and most of the 
metallicity tracers considered to date are outside of the crowded central regions of 
the galaxy (e.g. discussion by Magrini et al. 2017 and Patrick et al. 2017). While 
there is uncertainty as to the physical location of objects within the galaxy, 
the majority of the compact HII regions found in the present study are in 
the peripheral regions of H2 and H4, suggesting a physical connection with these 
structures. 

\noindent{4)} The spectra of some compact HII regions contain absorption lines of the NIR 
Ca triplet. This suggests that the areas in and/or around these HII regions harbor 
concentrations of luminous red stars when compared with the surrounding areas. 
The resolution elements that contain these HII regions 
thus contain stars with ages that span at least $\sim 10$ Myr.

\noindent{5)} The spectra of some unresolved objects suggest that their red/NIR light is 
dominated by early-type stars, while the spectra of others suggest that the light is 
dominated by RSGs or bright AGB stars. If these objects 
are star clusters then they will appear as diffuse 
structures in images with an angular resolution of $\sim 0.1$ arcsec. 

\noindent{6)} A subset of point sources have spectra that show 
CN absorption bands, suggesting that C stars make a large 
contribution to the light at red wavelengths. Clusters that contain 
very bright C stars are of interest for probing the star 
formation history (SFH) of NGC 55 during intermediate epochs 
(e.g. Maraston 2005). The depths of the CN bands are similar to those in the 
spectra of the LMC cluster NGC 1978, which has an age of 1.9 Gyr (Mucciarelli et 
al. 2007). While this similarity is likely a selection effect (Section 5), 
the discovery of clusters in NGC 55 with spectroscopic properties 
like NGC 1978 may not be unexpected given the various lines of evidence that 
support large scale star formation in NGC 55 a few Gyr in the past 
(e.g. discussion by Davidge 2018a, and below). 

	We close by discussing the evolution of NGC 55. 
The outer isophotes of NGC 55 have a lenticular shape, hinting that it 
is structurally distinct from other nearby late-type disk galaxies (Davidge 2018b). 
Kudritzki et al. (2016) discuss the chemical evolution of 
NGC 55 and conclude that there is the large-scale accretion and outflow of material. 
If a large outflow persists then NGC 55 may eventually shed its ISM, 
leaving a gas-poor intermediate mass disk galaxy. 
Barway et al. (2013) compare the stellar contents of intrinsically bright and 
faint S0 galaxies, and find that whereas brighter S0 galaxies are dominated by older 
populations, lower mass lenticular systems contain a significant young component. 
Evidence for recent star formation is seen in nearby low mass S0 galaxies. NGC 404 and 
NGC 5102 are nearby moderate mass, S0 galaxies that have relatively old disks 
(Williams et al. 2010; Davidge 2015). However, both have nuclei that contain 
young or intermediate age stars (Seth et al. 2010; Davidge 2015). 
There is also evidence for previous episodes of star formation 
near their centers (e.g. Kacharov et al. 2018; discussion by Seth et al. 2010), 
indicating that their central regions are not simple stellar systems. 
Some low mass S0 galaxies in the Virgo cluster also host 
centrally-concentrated star formation (Davidge 2018c).

	Barway et al. (2013) suggest that intrinsically bright S0s 
formed early-on, whereas fainter S0s are the result of secular processes. 
If disk galaxies are transformed into lenticular galaxies throughout the age of the 
Universe then transition objects should be seen today. That NGC 404 and NGC 5102 
are located within a few Mpc suggests that transition objects may not be rare. 
In fact, based on the lenticular shape of its outer isophotes, coupled with 
a star-forming history that suggests that there was a major episode of star formation 
$\sim 2$ Gyr in the past, we suggest that NGC 55 is such a transition galaxy. 

	Could the physical cause of the proposed on-going morphological transformation 
be an interaction with another galaxy? A tidal interaction could explain the asymmetric 
distribution of stars and gas in the NGC 55 disk. 
The space velocities of NGC 300 and NGC 55 suggest that they may have 
interacted 1 -- 2 Gyr in the past (Westmeier et al. 2013), and there is evidence 
for elevated levels of star formation in NGC 55 during intermediate epochs. Davidge (2018a) 
reviewed the SFH of NGC 55, and noted that the C star frequency is an order of magnitude 
higher than in the LMC and SMC (Battinelli \& Demers 2005). Given that 
the mean metallicity of NGC 55 is comparable to that in the LMC then this argues 
that a larger fraction of the stellar mass of NGC 55 formed during intermediate 
epochs than in the LMC. This is noteworthy as there are 
large peaks in the SFR of the LMC during the epochs in which the progenitors of 
C stars form (e.g. Rubele et al. 2012), and these have been attributed to interactions with 
the SMC and the Galaxy (e.g. Bekki \& Chiba 2005). Rubele et al. (2012) estimate 
that 15 -- 30\% of the stellar mass in the parts of the LMC they 
studied formed at this time. The specific SFR of NGC 55 during intermediate 
epochs would then have been even larger than that in the LMC. There is a need to produce a 
detailed SFH of NGC 55 to place further constraints on the amplitude and 
duration of star-forming episodes in NGC 55 during the time of the proposed 
interaction.

	Bekki \& Chiba (2005) point out that globular clusters would likely have
been produced during the violent star-forming episode in the LMC that was driven 
by interactions. If NGC 55 experienced a similar event then large compact star 
clusters may also have formed. While there are no known globular clusters associated with 
NGC 55, the proposed compact clusters that are thought to host the C 
stars found in the present study may be the consequence of large scale 
star formation during intermediate epochs.

	The notion that NGC 55 experienced a major 
interaction during intermediate epochs is not without problems. 
While NGC 300 is an obvious candidate for the perturbing system, the star-forming 
history of NGC 300 is similar to that of NGC 2403 (Kang et al. 2017), suggesting 
that if an interaction with NGC 55 did occur it did not have a major impact on the 
stellar content of NGC 300. This is surprising as the tilted ring model applied 
to NGC 55 by Westmeier et al. (2013) yields a rotation curve that is similar to that 
of NGC 300, indicating comparable masses -- niavely, both galaxies would then be expected 
to show signatures of an interaction if there was an encounter 
between them. Still,the C star frequencies of NGC 55 and NGC 300 
are comparable (Battinelli \& Demers 2005), suggesting similar SFRs during 
intermediate epochs.

	Another problem with NGC 300 as the perturbing 
system is the comparatively low metallicity of the 
extraplanar HII regions examined by Tullmann et al. (2003), which indicate that NGC 55 is 
accreting (or has accreted) gas from a source other than the disk of a moderately large 
spiral galaxy like NGC 300. This being said, Westmeier et al. (2013) suggest ESO294--010 as 
a possible system that could have interacted with NGC 55. An interaction may also have 
involved a now defunct gas-rich dwarf galaxy. If the orbit of such a companion 
was coplanar with the disk of NGC 55 then it might be the cause of the asymmetric 
nature of the NGC 55 disk. 

	External interactions may not be required to form 
a lenticular galaxy. Intermediate mass disk galaxies that 
evolve in isolation are subject to multiple episodes of bar formation 
and buckling, ultimately resulting in the formation of pseudo-bulges (e.g. Kwak et al. 
2017). If this occured in NGC 55 then the buckling of the bar would have presumably 
spurred a large episode of star formation during intermediate epochs 
and produced isophotes with a lenticular shape. However, such a purely 
secular process does not explain the asymmetric distribution of the NGC 55 disk.
Westmeier et al. (2013) suggest that ram pressure stripping may explain the asymmetric 
disk of NGC 55, although it is not clear how this could produce an extended 
tail of stellar material along the disk plane.

\acknowledgements{Thanks are extended to the anonymous referee for providing a 
prompt and helpful report.}


\parindent=0.0cm

\begin{references}

\reference{}Bacon, R., Accordo, M., Adjali, L., et al. 2010, Proc. SPIEE, 7735, 8

\reference{}Badenes, C., Maoz, D., Draine, B. T. 2010, MNRAS, 407, 1301

\reference{}Barnes, A. T., Longmore, S. N., Battersby, C., Bally, J., Kruijssen, J. M. D., Henshaw, J. D., \& Walker, D. L. 2017, MNRAS, 469, 2263

\reference{}Barway, S., Wadedekar, Y., Vaghmare, K., Kembhavi, A. K. 2013, MNRAS, 432, 430

\reference{}Battinelli, P., \& Demers, S. 2005, A\&A, 434, 657

\reference{}Bekki, K., \& Chiba, M. 2005, MNRAS, 356, 680

\reference{}Blair, W. P., Kirshner, R. P., \& Chevalier, R. A. 1981, ApJ, 247, 879

\reference{}Byler, N., Dalcanton, J. J., Conroy, C., \& Johnson, B. D. 2017, ApJ, 840, 44

\reference{}Carlos Reyes, R. E., Reyes Navarro, F. A., Melendez, J., Steiner, J., \& Elizalde, F. 2015, Rev Mex AA, 51, 135

\reference{}Castro, N., Urbaneja, M. A., Herrero, A., et al. 2012, A\&A, 542, A79

\reference{}Davidge, T. J. 1998, ApJ, 497, 650

\reference{}Davidge, T. J. 2005, ApJ, 622, 279

\reference{}Davidge, T. J. 2015, ApJ, 799, 97

\reference{}Davidge, T. J. 2018a, ApJ, 856, 129

\reference{}Davidge, T. J. 2018b, RNAAS, 2, A206

\reference{}Davidge, T. J. 2018c, AJ, 156, 233

\reference{}Davidge, T. J., Beck, T. L., \& McGregor, P. J. 2010, PASP, 112, 241

\reference{}De Marchi, G., Paresce, F., Panagia, N., et al. 2011, ApJ, 739, 27

\reference{}De Marchi, G., Panagia, N., \& Beccari, G. 2017, ApJ, 846, 110

\reference{}de Vaucouleurs, J., de Vaucouleurs, A., Corwin, H. G., Buta, R. J., Pateral, G., \& Fouque, P. 1991, The Third Reference Catalogue of Galaxies, Springer-Verlag: New York

\reference{}Engelbracht, C. W., Gordon, K. D., Bendo, G. J., et al. 2004, ApJS, 154, 248

\reference{}Espinoza, P., Selman, F. J., \& Melnick, J. 2009, A\&A, 501, 563

\reference{}Ferguson, A. M. N., Wyse, R. F. G., \& Gallagher, J. S. 1996, AJ, 112, 2567

\reference{}Gil de Paz, A., Boissier, S., Madore, B. F., et al. 2007, ApJS, 173, 185

\reference{}Graham, J. A. 1982, ApJ, 252, 474

\reference{}Graham, J. A., \& Lawrie, D. G. 1982, ApJ, 253, L73

\reference{}Hollyhead, K., Bastian, N., Adamo, A., Silva-Villa, E., Dale, J., Royon, J. E., \& Gazak, Z. 2015, MNRAS, 449, 1106

\reference{}Hook, I. M., Jorgensen, I., Allington-Smith, J. R., Davies, R. L., Metcalfe, N., Murowinski, R. G., \& Crampton, D. 2004, PASP, 116, 425

\reference{}Hoopes, C. G., Walterbos, R. A. M., \& Greenawalt, B. E. 1996, AJ, 112, 1429

\reference{}Jarrett, T. H., Chester, T., Cutri, R., Schneider, S. E., \& Huchra, J. P. 2003, AJ, 125, 525

\reference{}Kacharov, N., Neumayer, N., Seth, A. C., Cappellari, M., McDermid, R., Walcher, C. J., \& Torsten, B. 2018, MNRAS, 480, 1973

\reference{}Kang, X., Zhang, F., \& Chang, R. 2017, MNRAS, 469, 1636

\reference{}Karachentsev, I. D., Grebel, E. K., Sharina, M. E., et al. 2003, A\&A, 404, 93

\reference{}Kudritzki, R. P., Castro, N., and Urbaneja, M. A. et al. 2016, ApJ, 829, 70

\reference{}Leitherer, C., Schaerer, D., Goldader, J., et al. 1999, ApJS, 123, 3

\reference{}Long, K. S., Blair, W. P., Winkler, P. F. et al. 2010, ApJS, 187, 405

\reference{}Magrini, L., Goncalves, D. R., \& Vajgel, B. 2017, MNRAS, 464, 739

\reference{}Maraston, C. 2005, MNRAS, 362, 799

\reference{}Martin, D. C., Fanson, J., Schiminovich, D., et al. 2005, ApJ, 619, L1

\reference{}Mucciarelli, A., Ferraro, F. R., Origlia, L., \& Fussi Pecci, F. 2007, AJ, 133, 2053
 
\reference{}Murray, N. 2011, ApJ, 729, 133

\reference{}Otte, B., \& Dettmar, R.-J. 1999, A\&A, 343, 705

\reference{}Pagel, B. E. J., Edmunds, H. G., Fosbury, R. A. E., \& Webster, B. L. 1978, MNRAS, 184, 569

\reference{}Patrick, L. R., Evans, C. J., Davies, B., et al. 2017, MNRAS, 468, 492

\reference{}Pritchet, C. J., Richer, H. B., Schade, D., Crabtree, D., \& Yee, H. K. C. 1987, ApJ, 323, 79

\reference{}Reach, W. T., Megeath, S. T., Cohen, M. et al. 2005, PASP, 117, 978

\reference{}Reiter, M., Marengo, M., Hora, J. L., \& Fazio, G. G. 2015, MNRAS, 447, 3909

\reference{}Roth, M. M., Sandin, C., Kamann, S., et al. 2018, A\&A, 618, A3

\reference{}Rubele, S., Kerber, L., Girardi, L., et al. 2012, A\&A, 537, A106

\reference{}Seth, A. C., Cappellari, M., Neumayer, N., et al. 2010, ApJ, 714, 713

\reference{}Sheth, K., Regan, M., Hinz, J. L., et al. 2010, PASP, 122, 1397

\reference{}Skrutskie, M. F., Cutri, R. M., Stiening, R. et al. 2006, AJ, 131, 1163

\reference{}Stephens, A. W., Frogel, J. A., DePoy, D. L., et al. 2003, AJ, 125, 2473

\reference{}Stetson, P. B. 1987, PASP, 99, 191

\reference{}Stetson, P. B., \& Harris, W. E. 1988, AJ, 96, 909

\reference{}Sturch, L., \& Madore, B. 2012, BAAS, 2194, 1107

\reference{}Tikhonov, N. A., Galazutdinova, O. A., \& Drozdovsky, I. O. 2005, A\&A, 431, 127

\reference{}Toribio San Cipriano, L., Dominguez-Guzman, G., Esteban, C., et al. 2017, MNRAS, 467, 3759

\reference{}Tullmann, R., Rosa, M. R., Elwert, T., et al. 2003, A\&A, 412, 69

\reference{}Vilchez, J. M., \& Esteban, C. 1996, MNRAS, 280, 720

\reference{}Webster, B. L., \& Smith, M. G. 1983, MNRAS, 204, 743

\reference{}Werner, M., Roellig, T., Low, F., et al. 2004, ApJS, 154, 1

\reference{}Westmeier, T., Koribalski, B. S., \& Braun, R. 2013, MNRAS, 434, 3511

\reference{}Williams, B. F., Dalcanton, J. J., Gilbert, K. M., et al. 2010, ApJ, 716, 71

\end{references}
\end{document}